\documentclass[showpacs,preprintnumbers,amsmath,amssymb,nofootinbib,floatfix]{revtex4}

\usepackage[dvips]{graphicx}
\usepackage{dcolumn}
\usepackage{bm}

\def\mapgeq{\mathbin{\lower.3ex\hbox{$\buildrel>\over{\smash{\scriptstyle\sim}\vphantom{_x}}$}}}
\def\mapleq{\mathbin{\lower.3ex\hbox{$\buildrel<\over{\smash{\scriptstyle\sim}\vphantom{_x}}$}}}
\def\mapgeqeq{\mathbi{\lower.3ex\hbox{$\buildrel>\over{\smash{\scriptstyle\approx}\vphantom{_2}}$}}}
\def\mapleqeq{\mathbin{\lower.3ex\hbox{$\buildrel<\over{\smash{\scriptstyle\approx}\vphantom{_2}}$}}}
\mathchardef\hanaO="724F
\def\Journal#1#2#3#4{{#1} {\bf #2} (#4) #3}
\def\MPL{Mod. Phys. Lett. A}

\def\NPB{Nucl. Phys. B}

\def\NPSUPPL{Nucl. Phys. Proc. Suppl.}
\def\PLB{{Phys. Lett.} B}

\def\PLBOLD{Phys. Lett.}
\def\PRL{Phys. Rev. Lett.}
\def\RMP{Rev. Mod. Phys.}
\def\PRD{Phys. Rev. D}

\def\PTP{Prog. Theor. Phys.}
\def\JHEP{JHEP}

\def\EPJ{Euro. Phys. J. C}

\def\JETPUSSR{Sov. Phys. JETP}
\def\JETPUSSRLETT{Sov. Phys. JETP Letters}
\def\ZETP{Zh. Eksp. Teor. Fiz.}
\def\PismaZETP{Pis'ma Zh. Eksp. Teor. Fiz.}

\def\IJMP{Int. J. Mod. Phys. A}

\def\JPG{J. Phys. G}

\def\SCI{Science}
\def\APJ{Astrophys. J.}

\def\NJP{New. J. Phys.}
\def\PPNP{Prog. Part. Nucl. Phys.}

\def\Erratum{Erratum-ibid}

\begin{document}

\preprint{TOKAI-HEP/TH-0603}

\title{Correlation between Leptonic CP Violation and $\mu$-$\tau$ Symmetry Breaking}

\author{Teppei Baba}
\email{5asnm016@keyaki.cc.u-tokai.ac.jp}
\author{Masaki Yasu\`{e}}%
\email{yasue@keyaki.cc.u-tokai.ac.jp}
\affiliation{\vspace{3mm}%
\sl Department of Physics, Tokai University,\\
1117 Kitakaname, Hiratsuka, Kanagawa 259-1292, Japan\\
}

\date{December, 2006}

\begin{abstract}
Considering the $\mu$-$\tau$ symmetry, 
we discuss a direct linkage between phases of flavor neutrino masses and leptonic CP violation 
by determining three eigenvectors associated with ${\rm\bf M}=M^\dagger_\nu M_\nu$ for 
a complex flavor neutrino mass matrix $M_\nu$ in the flavor basis.  Since the Dirac CP violation is absent 
in the $\mu$-$\tau$ symmetric limit, leptonic CP violation is sensitive 
to the $\mu$-$\tau$ symmetry breaking, whose effect can be evaluated by perturbation. 
It is found that the Dirac phase ($\delta$) arises from the $\mu$-$\tau$ symmetry breaking part of 
${\rm\bf M}_{e\mu,e\tau}$ and an additional phase ($\rho$) is associated with 
the $\mu$-$\tau$ symmetric part of ${\rm\bf M}_{e\mu,e\tau}$, 
where ${\rm\bf M}_{ij}$ stands for an $ij$ matrix element ($i,j$=$e,\mu,\tau$). 
The phase $\rho$ is redundant and can be removed but leaves its 
effect in the Dirac CP violation characterized by $\sin (\delta  + \rho)$.  
The perturbative results suggest the exact formula of mixing parameters 
including that of $\delta$ and $\rho$, which turns out to be free from the effects of the redundant phases. 
As a result, it is generally shown that the maximal atmospheric neutrino mixing necessarily accompanies either 
$\sin\theta_{13}=0$ or $\cos(\delta+\rho)=0$, the latter of which indicates maximal CP violation, where 
$\theta_{13}$ is the $\nu_e$-$\nu_\tau$ mixing angle.
\end{abstract}

\maketitle
\section{\label{sec:1}Introduction}
Properties of neutrinos have been extensively studied by various experiments \cite{Sun,K2K,Reactor,Experiments}
 since the historical confirmation of the atmospheric neutrino oscillations by the Super-Kamiokande collaboration \cite{SK}.  
 For our understanding of hidden properties of the neutrinos  it is physically significant to observe leptonic 
CP violation in future neutrino experiments \cite{CPProposal} 
since there is no reason that prevents the appearance of CP violation in the lepton sector.  
Theoretically, leptonic CP violation is 
sensitive to phases \cite{CPViolationOrg} of the PMNS (Pontecorvo-Maki-Nakagawa-Sakata) unitary matrix $U_{PMNS}$ that converts the 
flavor neutrinos $\nu_{e,\mu,\tau}$ into the massive neutrinos $\nu_{1,2,3}$: $\nu_f = (U_{PMNS})_{fi}\nu_i$ 
($f$=$e,\mu,\tau$; $i$=1,2,3) 
\cite{PMNS}, where a neutrino mass matrix $M_\nu$ in the flavor basis is transformed into 
$U^T_{PMNS} M_\nu U_{PMNS}$=diag.($m_1, m_2, m_3$). CP effects 
in neutrino reactions are known to be characterized by the Jarlskog invariant ${\mathcal J}_{CP}$ proportional to 
$\sin\theta_{13} \sin\delta$ \cite{Jarlskog}, where $\theta_{13}$ and $\delta$, respectively, stand 
for the $\nu_e$-$\nu_\tau$ mixing angle and the CP violating Dirac phase.  However, these two quantities 
are not experimentally well known and the current data show the upper bound on $\theta_{13}$ 
\cite{NuData}:
\begin{eqnarray}
\sin ^2 \theta _{13}  = 0.9 
{\footnotesize
{\begin{array}{*{20}c}
   { + 2.3}  \\
   { - 0.9}  \\
\end{array}} \times 10^{-2},
}
\label{Eq:NuDataAngle13}
\end{eqnarray}
and no indication of the presence of $\delta$.  For a given value of $\theta_{13}$, CP violation 
becomes maximal if $\delta =\pi/2$ (modulo $\pi$).  It is of great importance to accumulate theoretical knowledge about 
$\theta_{13}$ and $\delta$.  We expect that it also provides useful information on the choice of CP phases in the leptogenesis \cite{leptogenesis} 
to create the baryon number in the Universe \cite{CP-Baryon}.

The leptonic CP violation is usually parameterized by phases in $U_{PMNS}$
 given by the particle data group \cite{PDG}, which we call $U^{PDG}_{PMNS}$, as 
 $U^{PDG}_{PMNS}\left(\delta\right)=U_\nu \left(\delta, 0, 0\right) K\left(\beta_1, \beta_2, \beta_3\right)$:
\begin{eqnarray}
U_\nu\left(\delta, 0, 0\right)&=&\left( \begin{array}{ccc}
  c_{12}c_{13} &  s_{12}c_{13}&  s_{13}e^{-i\delta}\\
  -c_{23}s_{12}-s_{23}c_{12}s_{13}e^{i\delta}
                                 &  c_{23}c_{12}-s_{23}s_{12}s_{13}e^{i\delta}
                                 &  s_{23}c_{13}\\
  s_{23}s_{12}-c_{23}c_{12}s_{13}e^{i\delta}
                                 &  -s_{23}c_{12}-c_{23}s_{12}s_{13}e^{i\delta}
                                 & c_{23}c_{13}\\
\end{array} \right),
\nonumber \\
K\left(\beta_1, \beta_2, \beta_3\right) &=& {\rm diag}(e^{i\beta_1}, e^{i\beta_2}, e^{i\beta_3}),
\label{Eq:U_nu}
\end{eqnarray}
for $c_{ij}=\cos\theta_{ij}$ and $s_{ij}=\sin\theta_{ij}$ ($i,j$=1,2,3), 
where $\theta_{ij}$ stand 
for three neutrino mixing angles. The Majorana CP violation phases are determined by two combinations 
of $\beta_{1,2,3}$ such as $\beta_i-\beta_3$ ($i$=1,2,3).  Since the knowledge on flavor neutrino masses 
$M_\nu$ contains all information of new physics of neutrinos, to speak about CP violations, 
it is useful to find direct linkage between phases in $M_\nu$ and 
CP violating Dirac and Majorana phases.  However, phases in $M_\nu$ are not uniquely determined 
because of the freedom of the redefinition of phases of the flavor neutrinos 
without affecting physical consequences and thus contain redundant phases. 
This freedom dictates other versions of $U_{PMNS}$ than Eq.(\ref{Eq:U_nu}). 
The true effect of the Dirac CP violation can only be 
discussed after obtaining $U_{PMNS}$ and is described in terms $(U_{PMNS})_{fi}$ through 
${\mathcal J}_{CP}$, which is free from the effect of the redefinition. 
There is a useful formula without referring to $U_{PMNS}$ to determine the Jarlskog invariant in terms of 
$M^\dagger_\nu M_\nu(\equiv {\rm\bf M})$, whose matrix element is denoted by ${\rm\bf M}_{ij}$ ($i,j$=$e,\mu,\tau$) 
\cite{JarlskogMass}: 
${\mathcal J}_{CP}={\rm Im}({\rm\bf M}_{e\mu}{\rm\bf M}_{\mu\tau}{\rm\bf M}_{\tau e})/(\Delta m^2_{12}\Delta m^2_{23}\Delta m^2_{31})$, 
where $\Delta m^2_{ij}=m^2_i-m^2_j$ ($i,j$=1,2,3).  
However, in any cases, it is not an easy task to find from ${\mathcal J}_{CP}$ which 
flavor neutrino mass of ${\rm\bf M}$ is mainly responsible for the Dirac CP violating phase.

We have advocated the use of the $\mu$-$\tau$ symmetry \cite{Nishiura,mu-tau,mu-tau0,mu-tau1,mu-tau2} 
to evaluate effects of leptonic CP violation \cite{MassTextureCP1,MassTextureCP10,MassTextureCP11,MassTextureCP2}. 
It is based 
on the observation that the $\mu$-$\tau$ symmetric texture gives $\sin\theta_{13}=0$ (or $\sin\theta_{12}=0$ 
\cite{Fuki}) giving no Dirac CP violation.\footnote{For $\sin\theta_{12}=0$, textures must contain a small parameter denoted by $\eta$ and 
yield $\tan 2\theta_{12}\sim \varepsilon /\eta$, 
which vanishes as $\varepsilon\rightarrow 0$ but stays ${\mathcal O}(1)$ for $\eta\sim \varepsilon$, where 
$\varepsilon$ stands for a $\mu$-$\tau$ symmetry breaking parameter. The Dirac CP violating phase becomes $\delta+\rho$, 
which vanishes as will be shown in Eq.(\ref{Eq:Mdagger-M-mixing_12case-CPphase}).} 
Therefore, leptonic CP violation is sensitive to how the $\mu$-$\tau$ symmetry is broken.  
For example, if the $\mu$-$\tau$ symmetry breaking part 
of $M_\nu$ is restricted to be pure imaginary and if its symmetric part is real, the Dirac CP violation 
becomes maximal for given mixing angles and the atmospheric neutrino mixing is also becomes maximal 
\cite{Maximal,MassTextureCP1,MassTextureCP10,MassTextureCP11}.  
This texture shows $M_{e\tau}$=$-\sigma M^\ast_{e\mu}$ ($\sigma = \pm 1$) and $M_{\tau\tau}$=$M^\ast_{\mu\mu}$, where 
$M_{ij}$ ($i,j$=$e,\mu,\tau$) stands for the $i$-$j$ matrix element of $M_\nu$.\footnote{The parameter $\sigma$ 
is so chosen to yield $\sin \theta_{23}=\sigma /\sqrt{2}$ in Eq.(\ref{Eq:U_nu}) after diagonal zing 
the $\mu$-$\tau$ symmetric $M_\nu$, where $M_{e\tau}$=$-\sigma M_{e\mu}$ and 
$M_{\tau\tau}$=$M_{\mu\mu}$ are satisfied. This assignment of $\sigma$ also appears in 
Eq.(\ref{Eq:Mnu-mutau-separation-2}).} 
It can be further shown that this  condition is extended to a more general one: 
$\vert M_{e\tau}\vert= \vert M_{e\mu}\vert$ and 
$\vert M_{\tau\tau}\vert = \vert M_{\mu\mu}\vert$ \cite{MassTextureCP11}.\footnote{The condition is in 
fact found to be a solution to 
$\vert M_{e\mu}\vert^2 -\vert M_{e\tau}\vert^2 =\vert M_{\tau\tau}\vert^2 -\vert M_{\mu\mu}\vert^2$ (corresponding to 
${\rm\bf M}_{\mu\mu }={\rm\bf M}_{\tau\tau }$),  
which is the relation satisfied by a set of flavor neutrino masses discussed in Ref.\cite{MassTextureCP11}.}

It is known that the $\mu$-$\tau$ symmetry required for neutrinos is badly broken by charged leptons. Since 
a pair of the charged lepton and neutrino forms an $SU(2)_L$-doublet, discussions based on the $\mu$-$\tau$ 
symmetry required for neutrinos may be useless.  However, it should be noted 
that to talk about the experimental results of neutrino mixings is almost equivalent to talk about 
theoretical results based on the approximate $\mu$-$\tau$ symmetry.  This is based on the good guiding 
principle to understand sources of small quantities in physics, which is the naturalness \cite{naturalness}.  
The naturalness dictates that the 
smallness of a certain physical quantity (such as $\sin^2\theta_{13}\ll 1$ and/or 
$\Delta m^2_\odot/\vert\Delta m^2_{atm}\vert\ll 1$) 
implies a certain protection symmetry (such as the $\mu$-$\tau$ symmetry).  Furthermore, 
the similarity between charged leptons and neutrinos loses direct linkage 
because charged leptons are Dirac particles but neutrinos can be Majorana particles. This significant 
difference yields sufficient power to construct various models based on the $\mu$-$\tau$ 
symmetry \cite{mu-tau0,mu-tau1,leptonic-mu-tau,recent_mu-tau-breaking}. 

In this article, along this line of thought, we discuss properties of the leptonic CP violation 
without entailing details of textures and examine constraints on various flavor neutrino masses to be 
compatible with the current neutrino oscillation data.  We focus on describing  
the CP violating Dirac phase as well as the mixing angles in terms of flavor neutrino masses as general as 
possible. As a result, $U_{PMNS}$ is completely expressed in terms of flavor neutrino masses and 
we understand which flavor neutrino masses mainly control which part of the 
leptonic CP violation \cite{YasueTalk}.  To construct $U_{PMNS}$, we must find three eigenvectors associated 
with ${\rm\bf M}$ \cite{Fuki,YasueTalk,EigenVectors}.
We first determine eigenvectors for the $\mu$-$\tau$ 
symmetric part of ${\rm\bf M}$. In the symmetric limit, the eigenvectors are found to generally contain a redundant
phase to be denoted by $\rho$ in the $\nu_e$-$\nu_\mu$ mixing, which serves as an additional CP 
violating Dirac phase after the $\mu$-$\tau$ symmetry breaking effects are included.  Next,
to see the $\mu$-$\tau$ symmetry breaking effects, we employ the usual perturbative 
analysis of ${\rm\bf M}$, where the $\mu$-$\tau$ symmetry breaking part of ${\rm\bf M}$ is treated as a 
perturbation.  This perturbation yields another phase to be denoted by $\gamma$ associated with 
the $\nu_\mu$-$\nu_\tau$ mixing to 
fill $U_{PMNS}$ in a consistent manner.  However, this phase $\gamma$ is completely removed without affecting 
CP violation. On the other hand, to remove the phase $\rho$ affects CP violation and results in $\rho+\delta$ as 
a physical CP violating Dirac phase. We keep $\rho$ in the PMNS unitary matrix to trace its effect.  
  Considering the perturbative results, we are led to general formula 
to express these mixings, and CP phases in terms of ${\rm\bf M}$ itself.  

Our formula are shown to indeed reproduce 
the perturbative results and to consistently take care of the effect of the redundant phases $\rho$ and $\gamma$. 
It turns out that $\delta$ and $\rho$ are determined to be
\begin{eqnarray}
&&
\delta = -{\arg}\left(
\sin\theta_{23} {\rm\bf M}^\prime_{e\mu} + \cos\theta_{23} {\rm\bf M}^\prime_{e\tau}
\right)
,
\label{Eq:Exact-delta-Intro}\\
&&
\rho = {\arg}\left(
\cos\theta_{23} {\rm\bf M}^\prime_{e\mu} - \sin\theta_{23} {\rm\bf M}^\prime_{e\tau}
\right),  
\label{Eq:Exact-rho-Intro}
\end{eqnarray}
where ${\rm\bf M}^\prime_{e\mu}=e^{i\gamma } {\rm\bf M}_{e\mu}$, and 
${\rm\bf M}^\prime_{e\tau}=e^{-i\gamma } {\rm\bf M}_{e\tau}$ are 
the redefined masses, which mean that the effect of $\gamma$ is absorbed by the 
redefinition of masses. If $\rho$ is removed, ${\rm\bf M}^\prime_{e\mu,e\tau}$ are 
shifted to ${\rm\bf M}^\prime_{e\mu}=e^{i(\gamma-\rho) } {\rm\bf M}_{e\mu}$, and 
${\rm\bf M}^\prime_{e\tau}=e^{-i(\gamma+\rho) } {\rm\bf M}_{e\tau}$, which give 
 $\delta+\rho$ in place of $\delta$ in Eq.(\ref{Eq:Exact-delta-Intro}) as expected. 
The atmospheric neutrino mixing angle is also determined and given by 
\begin{eqnarray}
&&
\theta _{23}  =  \sigma\frac{\pi}{4}+\frac{\theta + \phi}{2},
\nonumber\\
&&
\sin \theta  = \frac{\sigma\sin\theta_{13}\sin 2\theta_{12} \cos \left( \delta + \rho \right)\Delta m^2_\odot}
{2N},
\quad
\sin \phi  = \frac{\kappa\left({\rm\bf M}_{\mu\mu}-{\rm\bf M}_{\tau\tau}\right)}{2N},
\label{Eq:Exact-cos23-Intro}
\end{eqnarray}
where 
$\kappa = {\rm Re}( {\rm\bf M}^\prime_{\mu\tau})/\left| {\rm Re}({\rm\bf M}^\prime_{\mu\tau})\right|$, 
$N=\sqrt {{\rm Re} ^2 \left( {{\rm\bf M}^\prime_{\mu\tau}} \right) + ({\rm\bf M}_{\mu\mu}-{\rm\bf M}_{\tau\tau}) ^2/4 }$, 
and ${\rm\bf M}^\prime_{\mu\tau}=e^{ - 2i\gamma }{\rm\bf M}_{\mu\tau}$. The observed mass difference squared 
$\Delta m_ \odot^2$ is given by $\Delta m_ \odot^2=m^2_2-m^2_1(>0)$. 
For textures with $\sin\theta_{13}\neq 0$, the maximal atmospheric neutrino mixing is indeed provided by the maximal 
CP violating Dirac phase giving 
$\cos(\delta+\rho)=0$ as 
long as ${\rm\bf M}_{\mu\mu}={\rm\bf M}_{\tau\tau}$ is satisfied.  It should be noted that 
the combination of 
${\rm\bf M}_{e\mu}{\rm\bf M}_{\mu\tau}{\rm\bf M}_{\tau e}(={\rm\bf M}_{e\mu}{\rm\bf M}_{\mu\tau}{\rm\bf M}^\ast_{e\tau})$ 
appearing in ${\cal J}_{CP}$ 
becomes independent of the redefinition of the flavor neutrinos as expected because of 
${\rm\bf M}^\prime_{e\mu}{\rm\bf M}^\prime_{\mu\tau}{\rm\bf M}^{\prime\ast}_{e\tau}={\rm\bf M}_{e\mu}{\rm\bf M}_{\mu\tau}{\rm\bf M}^\ast_{e\tau}$ 
for ${\rm\bf M}^\prime_{e\mu}=e^{i(\gamma-\rho) } {\rm\bf M}_{e\mu}$, ${\rm\bf M}^\prime_{e\tau}=e^{-i(\gamma+\rho) } {\rm\bf M}_{e\tau}$ 
and ${\rm\bf M}^\prime_{\mu\tau}=e^{ - 2i\gamma }{\rm\bf M}_{\mu\tau}$. 

This paper is organized as follows. In the next section Sec.\ref{sec:2}, we discuss physical consequence of 
the $\mu$-$\tau$ symmetric part of ${\rm\bf M}$.  
The new phase $\rho$ is generally associated with ${\rm\bf M}_{e\mu,e\tau}$.  
In Sec.\ref{sec:3}, the discussions are further 
extended to include textures without the $\mu$-$\tau$ symmetry.  
We first calculate $U_{PMNS}$ composed of three eigenvectors by the 
perturbative method, which treats a $\mu$-$\tau$ symmetry breaking part of ${\rm\bf M}$ as a perturbation. 
We next derive a set of formula suggested by the results of the perturbation  
to calculate the mixing angles and phases, which can be applicable to any textures 
without the approximate $\mu$ - $\tau$ symmetry.  
In Sec.\ref{sec:4}, properties of neutrino oscillations are discussed by using our results given by the formula.  
To see the power of our formula, a simple neutrino mass matrix is analyzed.
The final section is devoted to summary and discussions. 

\section{\label{sec:2}$\mu$-$\tau$ Symmetry}
Let us begin with defining a neutrino mass matrix $M_\nu$ parameterized by\footnote{It is understood 
that the charged leptons and neutrinos are rotated, if necessary, to give diagonal charged-current 
interactions and to define the flavor neutrinos of $\nu_e$, $\nu_\mu$ and $\nu_\tau$.}
\begin{eqnarray}
&& M_\nu = \left( {\begin{array}{*{20}c}
	M_{ee} & M_{e\mu} & M_{e\tau}  \\
	M_{e\mu} & M_{\mu\mu} & M_{\mu\tau}  \\
	M_{e\tau} & M_{\mu\tau} & M_{\tau\tau}  \\
\end{array}} \right).
\label{Eq:NuMatrixEntries}
\end{eqnarray}
The $\mu$-$\tau$ symmetry can be defined by the invariance of the flavor neutrino mass terms in the 
lagrangian under the interchange of $\nu_\mu \leftrightarrow \nu_\tau$ or $\nu_\mu \leftrightarrow -\nu_\tau$.
  As a result, we obtain $M_{e\tau}=M_{e\mu}$ and $M_{\mu\mu}=M_{\tau\tau}$ for $\nu_\mu \leftrightarrow \nu_\tau$
 or $M_{e\tau}=-M_{e\mu}$ and $M_{\mu\mu}=M_{\tau\tau}$ for $\nu_\mu \leftrightarrow -\nu_\tau$.  We use 
the sign factor $\sigma=\pm 1$  to have $M_{e\tau}=-\sigma M_{e\mu}$ for the $\mu$-$\tau$ symmetric 
part under the interchange of $\nu_\mu\leftrightarrow -\sigma\nu_\tau$. 
If $M_\nu$ has CP phases, we have to use the Hermitian matrix ${\rm\bf M}$ to find eigenvectors of $M_\nu$, which 
can be divided into the $\mu$-$\tau$ symmetric part ${\rm\bf M}_{sym}$ and 
the $\mu$-$\tau$ symmetry breaking part ${\rm\bf M}_b $ as shown in the Appendix \ref{sec:Appendix1}.

The $\mu$-$\tau$ symmetric part ${\rm\bf M}_{sym}$ in Eq.(\ref{Eq:Mdagger-M}) is analyzed 
in this section.  We introduce a phase parameter $\rho$ to express 
\begin{eqnarray}
B_+= e^{i\rho}\vert B_+\vert, 
\label{Eq:Angle-rho}
\end{eqnarray}
leading to
\begin{eqnarray}
&&
{\rm\bf M}_{sym}  = \left( \begin{array}{*{20}c}
   A & e^{i\rho}\vert B_+\vert & - \sigma e^{i\rho}\vert B_+\vert \\
   e^{-i\rho}\vert B_+\vert & D_+ & E_+   \\
   - \sigma  e^{-i\rho}\vert B_+\vert & E_+ & D_+\\
\end{array} \right).
\label{Eq:Mdagger-M-sym-phase}
\end{eqnarray}
Three eigenvalues denoted by $\Lambda_\pm$ and $\Lambda$ associated with ${\rm\bf M}_{sym}$ are found to be:
\begin{eqnarray}
&&
\Lambda _ \pm   = D_+  - \sigma E_+ + \vert B_+ \vert X_ \pm,  
\quad
\Lambda  = D_+  + \sigma E_+,
\label{Eq:Mdagger-M-EigenValues}
\end{eqnarray}
where
\begin{eqnarray}
&&
X_ \pm   = \frac{{A  - D_+  + \sigma E_+  \pm \sqrt {\left( {A  - D_+ + \sigma E_+} \right)^2  + 8
\vert B_+\vert^2 } }}{{2\vert B_+\vert }}.
\label{Eq:Mdagger-M-X-pm}
\end{eqnarray}
The corresponding eigenvectors are calculated to be
\begin{eqnarray}
&&
\left| \Lambda_- \right\rangle  = N_- \left( {\begin{array}{*{20}c}
   -X_ -   \\
   -e^{-i\rho}  \\
    \sigma e^{-i\rho}  \\
\end{array}} \right),
\quad
\left| \Lambda_+ \right\rangle  = N_+ \left( {\begin{array}{*{20}c}
   X_ + e^{i\rho}  \\
   1  \\
   { - \sigma }  \\
\end{array}} \right),
\quad
\left| \Lambda \right\rangle  = \frac{1}{{\sqrt 2 }}\left( {\begin{array}{*{20}c}
   0  \\
   \sigma   \\
   1  \\
\end{array}} \right),
\label{Eq:Mdagger-M-EigenVectors}
\end{eqnarray}
respectively, for $\Lambda_-$, $\Lambda_+$ and $\Lambda$, where $N_\pm = \sqrt{2 + X^2_\pm}$. There is an 
ambiguity to have an overall phase in each eigenvector.  The orthogonality condition is obviously 
satisfied because of $X_+X_-=-2$.  

Considering the phase $\rho$ in Eq.(\ref{Eq:Mdagger-M-EigenVectors}), 
we parameterize the PMNS unitary matrix to be: 
$U^{(0)}_{PMNS}\left(\delta,\rho\right)=U_\nu \left(\delta, \rho, 0\right)K\left(\beta_1, \beta_2, \beta_3\right)$:
with 
\begin{eqnarray}
U_\nu \left(\delta, \rho, 0\right) &=&\left( {\begin{array}{*{20}c}
   {c_{12} c_{13} }  \\
   { - c_{23} s_{12} e^{ - i\rho }  - s_{23} c_{12} s_{13} e^{i\delta } }  \\
   {s_{23} s_{12} e^{ - i\rho }  - c_{23} c_{12} s_{13} e^{i\delta } }  \\
\end{array} \begin{array}{*{20}c}
   {s_{12} c_{13} e^{i\rho } }  \\
   {c_{23} c_{12}  - s_{23} s_{12} s_{13} e^{i\left(\delta+\rho\right) } }  \\
   { - s_{23} c_{12}  - c_{23} s_{12} s_{13} e^{i\left(\delta+\rho\right) } }  \\
\end{array} \begin{array}{*{20}c}
   {s_{13} e^{ - i\delta } }  \\
   {s_{23} c_{13} }  \\
   {c_{23} c_{13} }  \\
\end{array}} \right).
\label{Eq:Mdagger-M-PMNS-CP}
\end{eqnarray}
It should be 
noted that the Dirac CP violation is characterized by $\sin\theta_{13} \sin (\delta  + \rho)$ instead 
of $\sin\theta_{13} \sin \delta$ as indicated by ${\mathcal J}_{CP}$. It is obvious that ${\rm\bf M}_{sym}$ is equivalent to 
\begin{eqnarray}
&&
{\rm\bf M}^\prime_{sym}  = \left( \begin{array}{*{20}c}
   A & e^{-i\rho} B_+ & - \sigma e^{-i\rho} B_+ \\
   e^{i\rho} B^\ast_+ & D_+ & E_+   \\
   - \sigma  e^{i\rho} B^\ast_+ & E_+ & D_+\\
\end{array} \right)
=
\left( \begin{array}{*{20}c}
   A & \vert B_+\vert & - \sigma \vert B_+\vert \\
   \vert B_+\vert & D_+ & E_+   \\
   - \sigma  \vert B_+\vert & E_+ & D_+\\
\end{array} \right),
\label{Eq:Mdagger-M-sym-phase-equiv}
\end{eqnarray}
corresponding to Eq.(\ref{Eq:NuMatrixEntries-2}) with $\alpha_+=\rho$ and $\alpha_-=0$. Since 
${\rm\bf M}^\prime_{sym}$ yields $U^{PDG}_{PMNS}$ with $\delta=0$, the phase $\rho$ disappears and no CP violation exists. 
However, if there are $\mu$-$\tau$ symmetry breaking effects, CP violation is active.  
It is anticipated that the effect of $\rho$ remains in 
Dirac CP violation, where $\delta$ in $U^{PDG}_{PMNS}$ is replaced by $\delta+\rho$ at $U^{(0)}_{PMNS}$. 
To trace the effect of $\rho$ explicitly, we keep $\rho$ in the PMNS unitary matrix hereafter. 

As have been pointed out in Ref.\cite{Fuki}, there are two categories that characterize the effect from the $\mu$-$\tau$ 
symmetry.  In terms of the mixing angles, the $\mu$-$\tau$ symmetry requires either $\sin\theta_{13}=0$ or 
$\sin\theta_{12}=0$.  In terms of the eigenvalues of Eq.(\ref{Eq:Mdagger-M-EigenValues}), two categories depend on 
the order of $\vert\Lambda_\pm\vert$ and $\vert\Lambda\vert$, namely, on how these three eigenvalues are 
assigned to $m_{1,2,3}$. 
For $m_1 = \Lambda_-$, $m_2 = \Lambda_+$, and $m_3 = \Lambda$, 
if ${\rm\bf M}$ gives $\vert\Lambda_-\vert<\vert\Lambda_+\vert<\vert\Lambda\vert$ as 
the normal mass ordering or $\vert\Lambda\vert<\vert\Lambda_-\vert<\vert\Lambda_+\vert$ as the 
inverted mass ordering, we find $U_{PMNS}$ described by
\begin{eqnarray}
&&
\left( {\left| {\Lambda _ -  } \right\rangle ,\left| {\Lambda _ +  } \right\rangle ,\left| \Lambda  \right\rangle } \right)
=
\left( {\frac{1}{{\sqrt {2 + \left( {X_ -  } \right)^2 } }}\left( {\begin{array}{*{20}c}
   { - X_ -  }  \\
   { - e^{ - i\rho } }  \\
   \sigma e^{ - i\rho }   \\
\end{array}} \right)\frac{1}{{\sqrt {2 + \left( {X_ +  } \right)^2 } }}\left( {\begin{array}{*{20}c}
   {X_ + e^{i\rho }  }  \\
   1  \\
   { - \sigma }  \\
\end{array}} \right) \frac{1}{{\sqrt 2 }}\left( {\begin{array}{*{20}c}
   0  \\
   \sigma   \\
   1  \\
\end{array}} \right)} \right),
\label{Eq:Mdagger-M-U_PMNS_13case}
\end{eqnarray}
which yields
\begin{eqnarray}
&&
\tan 2\theta _{12}  = \frac{{2\sqrt 2 \vert B_+\vert }}{{D_+  - \sigma E_+ } - A},
\quad
\tan 2\theta _{23}  =\sigma,
\quad
\sin \theta _{13}  =0.
\label{Eq:Mdagger-M-mixing_13case}
\end{eqnarray}
On the other hand, for $m_1 = \Lambda_+$, $m_2 = \Lambda$, and $m_3 = \Lambda_-$, 
if the eigenvalues satisfy $\vert\Lambda_+\vert<\vert\Lambda\vert<\vert\Lambda_-\vert$
 as the normal mass ordering or $\vert\Lambda_-\vert<\vert\Lambda_+\vert<\vert\Lambda\vert$ as the 
inverted mass ordering, we find that $U_{PMNS}$ is described by
\begin{eqnarray}
&&
\left( {\left| {\Lambda _ +  } \right\rangle ,\left| \Lambda \right\rangle ,\left| \Lambda_-  \right\rangle } \right)
=
\left( {\frac{1}{{\sqrt {2 + \left( {X_ +  } \right)^2 } }}\left( {\begin{array}{*{20}c}
   { X_ +  }  \\
   { \sigma e^{ - i\rho } }  \\
   -e^{ - i\rho }   \\
\end{array}} \right)\frac{1}{{\sqrt 2 }}\left( {\begin{array}{*{20}c}
   {0  }  \\
   1  \\
   { \sigma }  \\
\end{array}} \right) \frac{1}{{\sqrt {2 + \left( {X_ -  } \right)^2 } }}\left( {\begin{array}{*{20}c}
   -X_-e^{i\rho}  \\
   -\sigma   \\
   1  \\
\end{array}} \right)} \right),
\label{Eq:Mdagger-M-U_PMNS_12case}
\end{eqnarray}
where $e^{-i\rho}$ is multiplied to $\vert \Lambda_\pm\rangle$ and the relative sign due to $\sigma$ 
is adjusted to match $U^{(0)}_{PMNS}$, which yield
\begin{eqnarray}
&&
\sin \theta _{12}  =0,
\quad
\tan 2\theta _{23}  = -\sigma,
\quad
\tan 2\theta _{13}  = \frac{{2\sqrt 2 \sigma \vert B_+\vert }}{{A  - D_+ + \sigma E_+ }}.
\label{Eq:Mdagger-M-mixing_12case}
\end{eqnarray}
By comparing Eq.(\ref{Eq:Mdagger-M-U_PMNS_12case}) with Eq.(\ref{Eq:Mdagger-M-PMNS-CP}), 
we obtain 
\begin{eqnarray}
&&
\delta=-\rho.
\label{Eq:Mdagger-M-mixing_12case-CPphase}
\end{eqnarray}
The Dirac CP violation is absent in both cases 
because of $\sin\theta_{13} \sin (\delta  + \rho) = 0$ satisfied by either $\sin\theta_{13}=0$ or $\delta+\rho$=0.

We have more convenient formula for masses and mixing angles as shown in the Appendix \ref{sec:Appendix2} 
with $\gamma = 0$. 
These formula yield the same results given by the eigenvectors and eigenvalues. Applying ${\rm\bf M}_{sym}$ to 
these relations, we find that
\begin{eqnarray}
&&
{\rm Re} \left( E_+ \right)\cos 2\theta _{23}  =  - s_{13} c_{\rho  + \delta }\left| X\right|,
\quad
s_{13} s_{\rho  + \delta }\left| X\right|=0,
\label{Eq:Mdagger-M-mixing_relations}
\end{eqnarray}
and
\begin{eqnarray}
&&
X = \frac{\left( c_{23} +\sigma s_{23} \right)B_+ }{{c_{13} }}.
\quad
Y= \left( s_{23}  - \sigma c_{23}\right) B_+.
\label{Eq:Mdagger-M-mixing_X}
\end{eqnarray}
 By using $\tan\theta_{13}=(\sqrt{1+\tan^2 2\theta_{13}}-1)/\tan 2\theta_{13}=\tan 2\theta_{13}(1+\cdots)\propto \vert Y\vert$
 from Eq.(\ref{Eq:Mdagger-ExactMixingAngles12-13})  and noticing that $\cos 2\theta _{23}= (c_{23} +\sigma s_{23})(c_{23} -\sigma s_{23})$, 
we find that  the solution 
to Eq.(\ref{Eq:Mdagger-M-mixing_relations}) turns out to be either
\begin{eqnarray}
&&
c_{23} = \sigma s_{23},
\label{Eq:Mdagger-M-mixing_13}
\end{eqnarray}
leading $\sin\theta_{13}=0$ with $\delta=0$, or
\begin{eqnarray}
&&
c_{23} = -\sigma s_{23},
\label{Eq:Mdagger-M-mixing_12}
\end{eqnarray}
leading $\sin\theta_{12}=0$ with $\delta=-\rho$ because of Eq.(\ref{Eq:Mdagger-M-mixing_X}) with 
$\tan 2\theta_{13}e^{-i\delta}\propto Y$ and $B_+\propto e^{i\rho}$.  

\section{\label{sec:3}$\mu$-$\tau$ Symmetry Breaking and CP Phases}
The $\mu$-$\tau$ symmetry breaking ${\rm\bf M}_b$ in Eq.(\ref{Eq:Mdagger-M}) is analyzed in this section.  
We choose conventional perturbative analysis with ${\rm\bf M}_b$ treated as a perturbation to 
find eigenvectors and eigenvalues associated with ${\rm\bf M}$.  An additional phase $\gamma$ is induced by 
$E_-$ in ${\rm\bf M}_b$ as a main source, which is to be removed by the redefinition of $\nu_{\mu,\tau}$. The 
inclusion of ${\rm\bf M}_b$ creates either $\sin\theta_{13}\neq 0$ or $\sin\theta_{12}\neq 0$.  However,
 the perturbative treatment only allows to induce tiny magnitude of $\sin\theta_{12,13}$.  For $\sin\theta_{12}\sim 0$,
 we cannot give the consistent result with the observation, which requires that $\sin^2 2\theta_{12}= {\mathcal O}(1)$, 
 and physical consequence of textures with $\sin\theta_{12}= 0$ in the $\mu$-$\tau$ symmetric limit 
 will be discussed in a separate article.

\subsection{\label{subsec:3-2-1}Perturbative Results}
The starting eigenvectors and eigenvalues are provided by
\begin{eqnarray}
&&
\left| 1^{(0)}\right.\rangle =
\frac{1}{\sqrt{2}}
\left( \begin{array}{*{20}c}
   \sqrt{2}c^{(0)}_{12}\\
   -s^{(0)}_{12}e^{ - i\rho }  \\
   \sigma s^{(0)}_{12}e^{ - i\rho }   \\
\end{array} \right),
\quad
\left| 2^{(0)}\right.\rangle =
\frac{1}{\sqrt{2}}
\left( \begin{array}{*{20}c}
   \sqrt{2}s^{(0)}_{12}e^{i\rho }\\
   c^{(0)}_{12}  \\
   -\sigma c^{(0)}_{12} \\
\end{array} \right),
\quad
\left| 3^{(0)}\right.\rangle =
\frac{1}{\sqrt{2}}
\left( \begin{array}{*{20}c}
   0\\
   \sigma  \\
   1 \\
\end{array} \right),
\label{Eq:Mdagger-M-vector13-0}
\end{eqnarray}
and, from Eq.(\ref{Eq:Mdagger-ExactMasses}), 
\begin{eqnarray}
&&
m_1^{(0)2}  = c^{(0) 2}_{12} A + s^{(0) 2}_{12} \left( {D_+  - \sigma E_+ } \right) - 2\sqrt 2 c^{(0)}_{12} s^{(0)}_{12} 
\left| {B_+ } \right|,
\nonumber\\
&&
m_2^{(0)2}  = s^{(0) 2}_{12} A + c^{(0) 2}_{12} \left( {D_+  - \sigma E_+ } \right) + 2\sqrt 2 c^{(0)}_{12} s^{(0)}_{12} 
\left| {B_+ } \right|,
\nonumber\\
&&
m_3^{(0)2}  = D_+  + \sigma E_+,
\label{Eq:Mdagger-M-mass13-0}
\end{eqnarray}
for textures with $\sin\theta_{13}=0$, where we have used the superscript (0) for mixing angles determined by 
Eq.(\ref{Eq:Mdagger-M-mixing_13case}).  From these results, we can find the first order results 
$\vert {n^{(1)} } \rangle$ and $m_n^{(1)2}$ ($n=1,2,3$) as follows:
\begin{eqnarray}
&&
\left| {1^{(1)} } \right\rangle  = a_{13}^{(1)} \left| {3^{(0)} } \right\rangle,
\quad
\left| {2^{(1)} } \right\rangle  = a_{23}^{(1)} \left| {3^{(0)} } \right\rangle,\quad
 \left| {3^{(1)} } \right\rangle  =  - \left( {a_{13}^{(1) * } \left| {1^{(0)} } \right\rangle  + a_
{23}^{(1) * } \left| {2^{(0)} } \right\rangle } \right),
\label{Eq:Mdagger-M-vector13-1}
\end{eqnarray}
where 
\begin{eqnarray}
&&
a_{13}^{(1)}  = \sigma \frac{{\sqrt 2 c^{(0)}_{12} B_-^ *   - s^{(0)}_{12} \left( {D_-  - i\sigma E_- } \right)e^
{ - i\rho } }}{{m_1^{(0)2}  - m_3^{(0)2} }},
\quad
a_{23}^{(1)}  = \sigma \frac{{\sqrt 2 s^{(0)}_{12} B_-^ *  e^{i\rho }  + c^{(0)}_{12} \left( {D_-  - i\sigma E_- }
 \right)}}{{m_2^{(0)2}  - m_3^{(0)2} }},
\label{Eq:Mdagger-M-vector13-1-coeff}
\end{eqnarray}
and
\begin{eqnarray}
&&
m_1^{(1)2}  = m_2^{(1)2}  = m_3^{(1)2}  = 0.
\label{Eq:Mdagger-M-mass13-1}
\end{eqnarray}
The three eigenvectors are now described by $\vert n \rangle = \vert n^{(0)} \rangle + \vert n^{(1)} \rangle$
 ($n=1,2,3$).  For instance, $\vert 3 \rangle$ is calculated to be:
\begin{eqnarray}
&&
\left| 3 \right\rangle \approx 
\frac{1}{{\sqrt 2 }}\left( {\begin{array}{*{20}c}
   {\sigma \frac{{2\left( {2 - R\cos 2\theta^{(0)} _{12} } \right)B_ -   + \sqrt 2 \sin 2\theta^{(0)} _{12} \left( {D_ -   + \sigma iE_ -  } \right)e^{i\rho } }}{{2\Delta m_{atm}^2 }}}  \\
   {\sigma \left( {1 + \frac{{\left( {2 + R\cos 2\theta^{(0)} _{12} } \right)\left( {D_ -   + \sigma iE_ -  } \right) + \sqrt 2 R\sin 2\theta^{(0)} _{12} B_ -  e^{ - i\rho } }}{{2\Delta m_{atm}^2 }}} \right)}  \\
   {1 - \frac{{\left( {2 + R\cos 2\theta^{(0)} _{12} } \right)\left( {D_ -   + \sigma iE_ -  } \right) + \sqrt 2 R\sin 2\theta^{(0)} _{12} B_ -  e^{ - i\rho } }}{{2\Delta m_{atm}^2 }}}  \\
\end{array}} \right),
\label{Eq:Mdagger-M-mass13-1st}
\end{eqnarray}
where $\Delta m_ {atm}^2=m^2_3-(m^2_1+m^2_2)/2$, and $\vert R\vert\ll 1$ 
for $R\equiv \Delta m^2_\odot/\Delta m^2_{atm}$
is used.

The PMNS unitary matrix can be constructed to be: ($\vert 1\rangle$, $\vert 2\rangle$, $\vert 3\rangle$).  By 
using Eq.(\ref{Eq:Mdagger-M-mass13-1st}), we observe that the corresponding part in $U^{(0)}_{PMNS}$,
 {\it i.e.} ($s_{13} e^{ - i\delta }, s_{23} c_{13}, c_{23} c_{13}$)$^T$, cannot be constructed from $\vert 3\rangle$
 because of the presence of the imaginary parts in all entries. 
It is found that a new phase $\gamma$ defined in 
\begin{eqnarray}
&&
P_\gamma = 
\left( {\begin{array}{*{20}c}
   1  \\
   0  \\
   0  \\
\end{array} \begin{array}{*{20}c}
   0 \\
   e^{i\gamma}  \\
   0 \\
\end{array} \begin{array}{*{20}c}
   0 \\
   0 \\
   e^{-i\gamma}\\
\end{array}} \right),
\label{Eq:Mdagger-M-PMNS-matrix-gamma}
\end{eqnarray}
is responsible for all phases in Eq.(\ref{Eq:Mdagger-M-mass13-1st}).\footnote{There may be another phase ($\tau$) associated with the $2-3$ rotation as 
$
\left( {\begin{array}{*{20}c}
    {\cos \theta _{23} } & {\sin \theta _{23} e^{i\tau } }  \\
    { - \sin \theta _{23} e^{ - i\tau } } & {\cos \theta _{23} }  \\
\end{array}} \right)
$ 
for $(\nu_\mu,\nu_\tau)$, which is not appropriate to describe Eq.(\ref{Eq:Mdagger-M-mass13-1st}) in the limit of $\vert \tau\vert\ll 1$. 
The coexistence of $\gamma$ and $\tau$ is mathematically irrelevant because 
three phases of the Dirac type are sufficient to determine the PMNS unitary matrix. The phase $\tau$ can be absorbed 
into $\rho$ and $\gamma$.  
It can be proved that we have $\rho^\prime = \rho  + \tau /2$ and $\gamma^\prime = \gamma  + \tau /2$ in place 
of $\rho$ and $\gamma$ in Eq.(\ref{Eq:Mdagger-Redifinition}) with $\delta^\prime$=$\delta+\rho+\tau$ so that 
the freedom of $\tau$ expressed in terms of $\delta^\prime$, $\rho^\prime$ and $\gamma^\prime$ becomes hidden.}
In fact, the PMNS unitary matrix $U_{PMNS}$ that 
diagonalizes ${\rm\bf M}_{sym}+{\rm\bf M}_b$ can take the form of
\begin{eqnarray}
&&
U_{PMNS}\left(\delta,\rho,\gamma\right) = P_\gamma U^{(0)}_{PMNS}\left(\delta,\rho\right)=U_\nu\left(\delta,\rho,\gamma\right)K\left(\beta_1, \beta_2, \beta_3\right),
\label{Eq:Mdagger-M-PMNS-CP-full}
\end{eqnarray}
where
\begin{eqnarray}
&&
U_\nu\left(\delta,\rho,\gamma\right) = P_\gamma U_\nu\left(\delta,\rho,0\right).
\label{Eq:Mdagger-M-PMNS-CP-gamma}
\end{eqnarray}
We can identify $\vert 3\rangle$ with the third row in $U_{PMNS}$ and find that
\begin{eqnarray}
&&
\left| 3 \right\rangle  \equiv \left( {\begin{array}{*{20}c}
   {s_{13} e^{ - i\delta } }  \\
   {s_{23} e^{i\gamma } }  \\
   {c_{23} e^{ - i\gamma } }  \\
\end{array}} \right) \approx \frac{1}{{\sqrt 2 }}\left( {\begin{array}{*{20}c}
   {\sqrt 2 s_{13} e^{ - i\delta } }  \\
   {\sigma \left( {1 - \Delta  + i\gamma } \right)}  \\
   {1 + \Delta  - i\gamma }  \\
\end{array}} \right).
\label{Eq:Mdagger-M-PMNS-CP-element}
\end{eqnarray}
This identification leads to
\begin{eqnarray}
&&
s_{13} e^{ - i\delta }  \approx 
\sigma \frac{{\sqrt 2 \left( {2 - R\cos 2\theta _{12} } \right)B_-  + R\sin 2\theta _{12} \left( {D_-  + \sigma iE_- } \right)e^{i\rho } }}{{2\Delta m_{atm}^2 }},
\nonumber\\
&&
\cos 2\theta_{23}  \left(\approx 2\Delta\right)\approx
- \frac{{\left( {2 + R\cos 2\theta _{12} } \right)D_-  + \sqrt 2 R\sin 2\theta _{12} {\rm Re} \left( {B_- e^{ - i\rho } } \right)}}{{\Delta m_{atm}^2 }},
\nonumber\\
&&
\gamma  \approx
\frac{{\left( {2 + R\cos 2\theta _{12} } \right)\sigma E_-  + \sqrt 2 R\sin 2\theta _{12} {\rm Im} \left( {B_- e^{ - i\rho } } \right)}}{{2\Delta m_{atm}^2 }},
\label{Eq:Mdagger-M-PMNS-CP-angles}
\end{eqnarray}
where $\theta_{12} \approx \theta^{(0)}_{12}$.  

The phase $\gamma$ turns out to consistently take care of other extra phases in $\vert 1\rangle$
 and $\vert 2\rangle$.
It is known that the phase $\gamma$ can be removed by appropriate redefinition of the neutrinos. 
Let $\nu_{flavor}=(\nu_e,\nu_\mu,\nu_\tau)^T$ and  $\nu_{mass}=(\nu_1,\nu_3,\nu_3)^T$. 
Because $\nu_{flavor} = P_\gamma U^{(0)}_{PMNS}\nu_{mass}$, new flavor 
neutrinos given by $\nu^\prime_{flavor}=P^{-1}_\gamma\nu_{flavor}$ 
are transformed by $U^{(0)}_{PMNS}$.  In this flavor base, we obtain that $-{\mathcal{L}_m}$
=$\nu^{\prime T}_{flavor}M^\prime_\nu\nu^\prime_{flavor}/2$ with $\nu^\prime_{flavor} = U^{(0)}_{PMNS}\nu_{mass}$ and
\begin{eqnarray}
&&
M^\prime_\nu = \left( {\begin{array}{*{20}c}
  M_{ee} & e^{i\gamma} M_{e\mu } & e^{-i\gamma} M_{e\tau } \\
  e^{i\gamma} M_{e\mu} & e^{2i\gamma} M_{\mu \mu } & M_{\mu \tau } \\
  e^{-i\gamma} M_{e\tau} & M_{\mu \tau } & e^{2i\gamma} M_{\tau \tau } \\
\end{array}} \right),
\label{Eq:NewNuMatrixEntries}
\end{eqnarray}
which is nothing but Eq.(\ref{Eq:NuMatrixEntries-2}) with $\alpha_-=-\gamma$ and $\alpha_{e,+}=0$.  
The phase $\gamma$ is now absent in the PMNS unitary matrix for $M^\prime_\nu $.

If the $\mu$-$\tau$ symmetry breaking terms have similar strengths, we expect that 
$\left| B_-\right| \sim \left| D_-\right| \sim \left| E_-\right|$.  
If this is the case, Eq.(\ref{Eq:Mdagger-M-PMNS-CP-angles}) can be further reduced to
\begin{eqnarray}
&&
s_{13} e^{ - i\delta }  \approx 
 \frac{\sqrt 2\sigma B_-}{\Delta m_{atm}^2},
\quad
\cos 2\theta_{23}  \approx
 - \frac{2D_-}{\Delta m_{atm}^2},
\quad
\gamma  \approx
\frac{2\sigma E_-}{2\Delta m_{atm}^2},
\label{Eq:Mdagger-M-PMNS-CP-angles-simple}
\end{eqnarray}
because of $R\ll 1$.  Considering Eq.(\ref{Eq:Mdagger-M-PMNS-CP-angles-simple}) and Eq.(\ref{Eq:Angle-rho}) 
for $\rho$, we observe that 
\begin{itemize}
\item the $\mu$-$\tau$ symmetric $B_+$ is the main source of $\rho$,
\item the $\mu$-$\tau$ symmetry breaking $B_-$ is the main source of $\delta$,
\item the $\mu$-$\tau$ symmetry breaking $D_-$ is the main source of $\cos 2\theta_{23}$,
\item the $\mu$-$\tau$ symmetry breaking $E_-$ is the main source of $\gamma$.
\end{itemize}
These are our main results found in textures with the approximate $\mu$-$\tau$ symmetry.

\subsection{\label{subsec:3-2-2}Exact Results}
As suggested from the subsection \ref{subsec:3-2-1}, we may employ 
$M^\prime_\nu$ of Eq.(\ref{Eq:NewNuMatrixEntries}) and $U_{PMNS}$ of Eq.(\ref{Eq:Mdagger-M-PMNS-CP-full}).  
The direct calculation from $U^\dagger_{PMNS}{\rm\bf M}U_{PMNS}$=diag.($m^2_1, m^2_2, m^2_3$) yields three 
constraints and three masses expressed in terms of ${\rm\bf M}_{ij}$ ($i,j=e,\mu,\tau$) just corresponding 
to Eqs.(\ref{Eq:Mdagger-ExactMixingAngles12-13}), (\ref{Eq:Mdagger-ExactMixingAngles23}) and (\ref{Eq:Mdagger-ExactMasses}).  
Let us define new flavor neutrino masses and a new Dirac phase $\delta^\prime$: 
\begin{eqnarray}
&&
B' = e^{i\left( {\gamma  - \rho } \right)} B,
\quad
C' = e^{ - i\left( {\gamma  + \rho } \right)} C,
\quad
E' = e^{ - 2i\gamma } E,
\nonumber \\
&&
X^\prime = e^{-i\rho}X,
\quad
Y^\prime = e^{-i\rho}Y,
\quad
\delta ' = \delta  + \rho,
\label{Eq:Mdagger-Redifinition}
\end{eqnarray}
which are induced by the appropriate redefinition of the neutrinos with $\alpha_+=\rho$ and $\alpha_-=-\gamma$ 
in Eq.(\ref{Eq:NuMatrixEntries-2}).  In another words, it is equivalent to show that
\begin{eqnarray}
&&
\nu_{flavor} = U_{PMNS}\left(\delta,\rho,\gamma\right)\nu_{mass}
=U_\nu\left(\delta,\rho,\gamma\right)K\left(\beta_1,\beta_2,\beta_3\right) \nu_{mass},
\nonumber\\
&&
\nu^\prime_{flavor}
=
U^{PDG}_{PMNS}\left(\delta^\prime\right) \nu_{mass}
=U_\nu\left(\delta^\prime,0,0\right)K\left(\beta_1-\rho,\beta_2,\beta_3\right) \nu_{mass},
\label{Eq:UnitaryMatrix-1}
\end{eqnarray}
as in  Eq.(\ref{Eq:NuRedifinition}) with $\alpha_e = -\rho$. 
 It is obvious that 
$BEC^\ast$ (=${\rm\bf M}_{e\mu}{\rm\bf M}_{\mu\tau}{\rm\bf M}_{\tau e}$) = 
$B^\prime E^\prime C^{\prime\ast}$ (=${\rm\bf M}^\prime_{e\mu}{\rm\bf M}^\prime_{\mu\tau}{\rm\bf M}^\prime_{\tau e}$) , which reassures that ${\mathcal J}_{CP}$ is a weak-base-independent quantity. 
The formula in the Appendix \ref{sec:Appendix2}  
are a little bit modified and relations to be modified are expressed in terms of the new masses and phase as follows:
\begin{eqnarray}
&&
\tan 2\theta _{12}  = \frac{{2X^\prime}}{{\Lambda _2  - \Lambda_1 }},
\quad
\tan 2\theta _{13}e^{-i\delta^\prime}  = \frac{2Y^\prime}{{\Lambda _3  - A}},
\nonumber\\
&&
{\rm Re} \left( E^\prime \right)\cos 2\theta _{23}  + D_-\sin 2\theta _{23}
  + i{\rm Im} \left(E^\prime \right) =  - s_{13} e^{- i\delta^\prime} X^{\prime\ast},
\nonumber\\
&&
X^\prime= \frac{{c_{23}B^\prime - s_{23}C^\prime}}{{c_{13} }} (={\rm real}),
\quad
Y^\prime = s_{23} B^\prime + c_{23} C^\prime,
\nonumber\\
&&
\Lambda _2  = c_{23}^2 D + s_{23}^2 F - 2s_{23} c_{23} {\rm Re} \left(E^\prime \right),
\quad
\Lambda _3  = s_{23}^2 D + c_{23}^2 F + 2s_{23} c_{23} {\rm Re} \left(E^\prime \right).
\label{Eq:Mdagger-Redinied}
\end{eqnarray}
These relations together with other relations in Eqs.(\ref{Eq:Mdagger-ExactMasses}) and 
(\ref{Eq:Mdagger-MassParameters}) 
guarantee that diagonal masses are obtained by $U^{PDG}_{PMNS}$, which does not 
contain $\rho$ and $\gamma$.
To determine $\rho$ and $\delta$ can be rephrased as follows \cite{GeneralCP}:
\begin{eqnarray}
&&
\tan\theta_{23} 
= 
\frac{{\rm Im}(e^{-i\left( {\rho -\gamma} \right)}B)}{{\rm Im}(e^{-i\left( {\rho+\gamma } \right)} C)}
=
-
\frac{{\rm Im}(e^{i\left( {\delta -\gamma} \right)} C)}{{\rm Im}(e^{i\left( {\delta+\gamma } \right)}B)},
\label{Eq:Mdagger-Redinied-23}
\end{eqnarray}
owing to the absence of the imaginary part of $X^\prime$ and $e^{i\delta^\prime}Y^\prime$.  
Our formula can be used to examine textures 
with $\sin\theta_{12} \rightarrow 0$ in the $\mu$-$\tau$ symmetric limit.

From the relations, we find that
\begin{itemize}
\item the CP phase $\rho$ is the phase of the flavor neutrino mass: $c_{23} e^{i\gamma} B - s_{23} e^{-i\gamma}$ as
\begin{eqnarray}
&&
\rho = {\arg}\left(c_{23} e^{i\gamma} B - s_{23} e^{-i\gamma} C\right),
\label{Eq:Exact-rho}
\end{eqnarray}
\item the CP phase $\delta$ is the phase of the flavor neutrino mass: $s_{23} e^{i\gamma } B + c_{23} e^{ - i\gamma } C$ as
\begin{eqnarray}
&&
\delta = -{\arg}\left(s_{23} e^{i\gamma } B + c_{23} e^{ - i\gamma } C\right),
\label{Eq:Exact-delta}
\end{eqnarray}
\item the real part of Eq.(\ref{Eq:Mdagger-ExactMixingAngles23})
\begin{eqnarray}
&&
{\rm Re} \left( {e^{ - 2i\gamma } E} \right)\cos 2\theta _{23}  + D_ -  \sin 2\theta _{23} 
=  -s_{13} \cos \left(\rho  + \delta \right) e^{i\rho} X^\ast\left( \equiv -x\right),
\label{Eq:Exact-Angle23}
\end{eqnarray}
determines $\cos 2\theta_{23}$, which is given by
\begin{eqnarray}
&&
\cos 2\theta _{23}  =  -  \frac{{\kappa\sigma D_ -\sqrt {{\rm Re} ^2 \left( {e^{ - 2i\gamma } E} \right) + D_ - ^2  - x^2 }   + x{\rm Re} \left( {e^{ - 2i\gamma } E} \right)}}{{{\rm Re} ^2 \left( {e^{ - 2i\gamma } E} \right) + D_ - ^2 }}
=  \cos \left( \sigma\frac{\pi}{2}+\theta + \phi \right),
\nonumber\\
&&
\cos \theta  = \sqrt {\frac{{\rm Re} ^2 \left( {e^{ - 2i\gamma } E} \right) + D_ - ^2  - x^2 }{{\rm Re} ^2 \left( {e^{ - 2i\gamma } E} \right) + D_ - ^2  }},
\quad
\sin \theta  = \frac{\sigma x}{\sqrt {{\rm Re} ^2 \left( {e^{ - 2i\gamma } E} \right) + D_ - ^2 } },
\nonumber\\
&&
\cos \phi  = \frac{{\rm Re} \left( {e^{ - 2i\gamma } E} \right)}{\sqrt {{\rm Re} ^2 \left( {e^{ - 2i\gamma } E} \right) + D_ - ^2 } },
\quad
\sin \phi  = \frac{\kappa D_-}{\sqrt {{\rm Re} ^2 \left( {e^{ - 2i\gamma } E} \right) + D_ - ^2 } },
\label{Eq:Exact-cos23-solution}
\end{eqnarray}
where $\kappa$ is the sign of ${\rm Re}( e^{ - 2i\gamma }E)$, from which we obtain that 
$\theta _{23} = \sigma\pi/4 + (\theta + \phi)/2$.
\item the imaginary part of Eq.(\ref{Eq:Mdagger-ExactMixingAngles23}) 
\begin{eqnarray}
&&
\cos 2\gamma {\rm Im} \left( E \right) - \sin 2\gamma {\rm Re}\left( E \right) = s_{13} \sin \left( {\rho  + \delta } \right)e^{i\rho} X^\ast\left( \equiv x^\prime\right),
\label{Eq:Exact-gamma}
\end{eqnarray}
determines $\gamma$, which is given by
\begin{eqnarray}
&&
{\sin 2\gamma  = \frac{{ \kappa^\prime {\rm Im}\left( E \right)\sqrt {\left| E \right|^2  - x^{\prime 2} } - x^\prime {\rm Re} \left( E \right)}}{{\left| E \right|^2 }} 
=  \sin \left( \phi^\prime-\theta^\prime \right)}.
\nonumber\\
&&
\cos \theta^\prime  = \frac{\sqrt {\left| E \right|^2  - x^{\prime 2} }}{{\left| E \right|}},
\quad
\sin \theta^\prime  = \frac{x^\prime}{{\left| E \right|}}
\nonumber\\
&&
\cos \phi^\prime  = \frac{{\rm Re} \left( E \right)}{{\left| E \right|}},
\quad
\sin \phi^\prime  = \frac{\kappa^\prime{\left| {{\rm Im} \left( E \right)} \right|}}{{\left| E \right|}},
\label{Eq:Exact-gamma-solution}
\end{eqnarray}
where $\kappa^\prime$ is the sign of ${\rm Re}( E)$, from which we obtain that 
$\gamma = (\phi^\prime-\theta^\prime)/2 $.
\end{itemize}
These exact results can be used to examine any textures.  
If textures are 
approximately $\mu$-$\tau$ symmetric, the perturbative result can be shown to be reproduced. 

To see that the perturbative result Eq.(\ref{Eq:Mdagger-M-PMNS-CP-angles}) is reproduced, we 
introduce the approximation due to $\vert\gamma\vert\ll 1$, $\vert\Delta\vert \ll 1$ and 
$\sin^2\theta_{13}\ll 1$.  Note that $\vert\gamma\vert\ll 1$ in Eq.(\ref{Eq:Exact-gamma-solution}) 
practically requires the suppression of 
${\rm Im}(E)$.  Using $\vert\gamma\vert\ll 1$,  
and ${\rm Re}\left(e^{ - 2i\gamma }  E \right)\approx \sigma \Delta m^2_{atm}/2$ together with 
$X = e^{i\rho}{{\sin 2\theta _{12} \Delta m_ \odot ^2 }}/{2}$ in Eq.(\ref{Eq:Mdagger-M-X}), 
it is readily found that
\begin{eqnarray}
&&
\cos 2\theta _{23} ( \approx 2\Delta) \approx 
 - \frac{{\left( {2 + R\cos 2\theta _{12} } \right)D_ -   + \sigma s_{13} \cos \left( {\rho  + \delta } \right)\sin 2\theta _{12} \Delta m_ \odot ^2 }}{{\Delta m_{atm}^2 }},
\nonumber\\
&&
\gamma  \approx 
\sigma \frac{{\left( {2 + R\cos 2\theta _{12} } \right) {\rm Im} \left( E \right) - s_{13} \sin \left( {\rho  + \delta } \right)\sin 2\theta _{12} \Delta m_ \odot ^2 }}{{2\Delta m_{atm}^2 }},
\label{Eq:Angle23-gamma-approx}
\end{eqnarray}
form Eqs.(\ref{Eq:Exact-cos23-solution}) and (\ref{Eq:Exact-gamma-solution}), 
which coincide with $\cos 2\theta_{23}$ and $\gamma$ in Eq.(\ref{Eq:Mdagger-M-PMNS-CP-angles}) because of 
$\sqrt 2 B_-  \approx \sigma s_{13} e^{ - i\delta } \Delta m_{atm}^2$. 
The mixing angle $\tan 2\theta_{13}$ in Eq.(\ref{Eq:Mdagger-ExactMixingAngles12-13}) with 
$\Lambda _3  - A \approx ( 2 + R\cos 2\theta _{12})\Delta m_{atm}^2/2$ becomes
\begin{eqnarray}
\tan 2\theta_{13}e^{-i\delta} 
&\approx&
{4\sqrt 2 \sigma \frac{{B_ - +\left( {i\gamma - \Delta } \right)B_+ }}{{\left( {2 + R\cos 2\theta _{12} } \right)\Delta m_{atm}^2 }}}
\nonumber\\
&\approx&
\sigma \frac{{\sqrt 2 \left( {2 - R\cos 2\theta _{12} } \right)} B_-
  + \left( {i\gamma  - \Delta } \right)
e^{i\rho } \sin 2\theta _{12} \Delta m_ \odot ^2 }{\Delta m_{atm}^2},
\label{Eq:Mdagger-M-13}
\end{eqnarray}
where Eq.(\ref{Eq:Mdagger-M-X}) is used to give $B_ +  \approx e^{ i\rho } \sin 2\theta _{12} \Delta m_ \odot ^2 / {2\sqrt 2 }$.
Inserting Eq.(\ref{Eq:Angle23-gamma-approx}) into Eq.(\ref{Eq:Mdagger-M-13}) and ignoring terms of ${\mathcal O}(R^2)$, 
Eq.(\ref{Eq:Mdagger-M-13}) coincides with the perturbative result Eq.(\ref{Eq:Mdagger-M-PMNS-CP-angles}).
The conditions of 
$\vert\cos 2\theta_{23}\vert\ll 1$, $\vert\gamma\vert\ll 1$, and 
$\vert\sin\theta_{13}\vert\ll 1$  are, respectively, 
satisfied if $\left| D_-\right| \ll \left| \Delta m^2_{atm}\right|$, $\left| {\rm Im}(E) \right| \ll \left| \Delta m^2_{atm}\right|$ 
and $\left| B_- \right| \ll \left| \Delta m^2_{atm}\right|$.  

Once again, predictions of  
the observed quantities such as neutrino masses and mixing angles are 
independent of $\gamma$ and $\rho$ because they depend on the 
modified neutrino masses in Eq.(\ref{Eq:Mdagger-Redifinition}), where $\rho$ and $\gamma$ are 
hidden by the definition of the masses.  However, it should be noted that $\gamma$ supplies $\mu$-$\tau$ 
symmetric breaking effects because the rotation induced by $\gamma$ is of the $\mu$-$\tau$ symmetry 
breaking type.  In approximately $\mu$-$\tau$ symmetric textures, $\vert\gamma\vert\ll 1$ should 
be satisfied.

\section{\label{sec:4}Properties of Neutrino Oscillations}
The mixing angles and associated phases are expressed in terms of the flavor neutrino masses. 
The CP violating Dirac phase $\delta^\prime=\delta+\rho$ is determined as 
$\delta = -\arg (Y)$ and $\rho = \arg (X)$, where
\begin{eqnarray}
&&
X = c_{23} e^{i\gamma} {\rm\bf M}_{e\mu } - s_{23} e^{-i\gamma} {\rm\bf M}_{e\tau },
\nonumber\\
&&
Y = s_{23} e^{i\gamma} {\rm\bf M}_{e\mu } + c_{23} e^{-i\gamma} {\rm\bf M}_{e\tau },
\label{Eq:SummaryPhases-XY}
\end{eqnarray}
which also controls the mixing angles as in Eq.(\ref{Eq:Mdagger-ExactMixingAngles12-13}) given by 
\begin{eqnarray}
&&
\sin 2\theta_{12}e^{i\rho} = \frac{2 X}{\Delta m^2_\odot},
\quad
\tan 2\theta_{13}e^{-i\delta} \approx \frac{2 Y}{\Delta m^2_{atm}},
\label{Eq:SummaryPhases-tan13}
\end{eqnarray}
where $\Lambda_2-\Lambda_1=\cos 2\theta_{12}\Delta m^2_\odot$ and $\Lambda_3-A\approx \Delta m^2_{atm}$ are used. 
To obtain $\sin^2 \theta_{13} \mapleq 0.01$ and 
$\tan 2\theta_{12}={\mathcal O}(1)$ calls for two constraints: 
1) $\left| X\right| \sim \Delta m^2_\odot$ and 
 2) $\left| Y\right| \mapleq 0.1 \left| \Delta m^2_{atm}\right|$.  If there is an approximate $\mu$-$\tau$ symmetry, 
$X\approx \sqrt 2\left( B_++B_-\Delta\right)$ and $Y\approx \sqrt 2 \left(B_-+B_+\Delta\right)$ are 
satisfied.  
Since the smallness of $\left| B_-\right|$ and $\Delta$ is a result of the approximate $\mu$-$\tau$ symmetry, 
 the second constraint is naturally expected.   
 However, the smallness of $\left| X\right|$ needs that of $\vert B_+\vert$ as an additional requirement. 

The deviation of the atmospheric neutrino mixing from its maximal one is estimated to be:
\begin{eqnarray}
&&
\theta _{23} = \sigma\frac{\pi}{4}+\frac{\theta  + \phi}{2},
\label{Eq:SummaryPhases-cos23}
\end{eqnarray}
where $\theta$, and $\phi$ are defined in Eq.(\ref{Eq:Exact-cos23-solution}). We observe 
that 
the phenomenological constraint that the atmospheric neutrino mixing is almost maximal is 
well satisfied if ${\rm\bf M}_{\mu\mu }\sim{\rm\bf M}_{\tau\tau }$ for $\phi\ll 1$ 
because $\theta\ll 1$ is always satisfied by 
$\left| X\right|= \Delta m^2_\odot/2\sin 2\theta_{12}$,
and 
${\rm Re}\left(e^{ - 2i\gamma } {\rm\bf M}_{\mu\tau} \right) \approx \sigma \Delta m_{atm}^2/2$ 
as in Eq.(\ref{Eq:Mdagger-M-MassRelation}).  
This constraint ${\rm\bf M}_{\mu\mu }\sim{\rm\bf M}_{\tau\tau }$ 
supplemented by $\vert Y\vert \sim 0$ 
indicates the presence of the approximate $\mu$-$\tau$ symmetry 
if the additional constraint $\vert \gamma \vert \ll 1$ leading to 
the suppression of $\vert {\rm Im}({\rm\bf M}_{\mu\tau })\vert$ 
is satisfied. 

We can also argue how the mass hierarchy of 
  $\vert \Delta m^2_{atm}\vert\gg \Delta m^2_\odot$
 constrains the magnitude of ${\rm\bf M}_{ee }$, ${\rm\bf M}_{\mu\mu }+{\rm\bf M}_{\tau\tau }$ and 
 ${\rm\bf M}_{\mu\tau }$ from Eq.(\ref{Eq:Mdagger-M-MassRelation}).  The results can be summarized as follows:
\begin{itemize}
\item for the normal mass hierarchy with $\vert\Delta m^2_{atm}\vert\gg \sum m^2_\odot$, 
${\rm\bf M}_{\mu\mu }+{\rm\bf M}_{\tau\tau }   \approx 2\sigma {\rm Re}( {e^{ - 2i\gamma } {\rm\bf M}_{\mu\tau }}) \approx {\Delta m_{atm}^2 } \gg 2{\rm\bf M}_{ee }$ 
are required,
\item for the inverted mass hierarchy ($\Lambda _3\ll \vert\Delta m^2_{atm}\vert$) and  
the degenerate mass pattern ($\Lambda _3\sim \vert\Delta m^2_{atm}\vert$ with $m^2_1\sim m^2_2\sim m^2_3\sim \vert\Delta m^2_{atm}\vert$), 
both with $2\vert\Delta m^2_{atm}\vert\approx \sum m^2_\odot$, 
$3{\rm\bf M}_{ee} \approx {\rm\bf M}_{\mu\mu }+{\rm\bf M}_{\tau\tau }   \approx  6\sigma {\rm Re} ( {e^{ - 2i\gamma } {\rm\bf M}_{\mu\tau }})\approx 3\Delta m_{atm}^2$ 
are required, and
\item for the degenerate mass pattern($\sum m^2_\odot \gg \vert\Delta m^2_{atm}\vert$ with $m^2_1\sim m^2_2\sim m^2_3\gg \vert\Delta m^2_{atm}\vert$), 
$2{\rm\bf M}_{ee } \approx {\rm\bf M}_{\mu\mu }+{\rm\bf M}_{\tau\tau }  \approx {\sum m_ \odot ^2 } \gg 2\sigma {\rm Re} ( {e^{ - 2i\gamma } {\rm\bf M}_{\mu\tau }} )$ 
are required.
\end{itemize}
The set of these constraints serves as a useful guide to search for specific textures.  

As an example, let us consider the simplest mass matrix discussed in Ref.\cite{Example}, which is 
\begin{eqnarray}
&&
M_\nu  = \frac{m_0}{2}\left( \begin{array}{*{20}c}
   a\varepsilon^2 & b\varepsilon & - \sigma b\varepsilon e^{i\alpha}  \\
   b\varepsilon & 1+\varepsilon & \sigma   \\
  - \sigma  b\varepsilon e^{i\alpha} & \sigma & 1+\varepsilon\\
\end{array} \right),
\label{Eq:Example-mass-matrix}
\end{eqnarray}
in our assignment, where $a,b$ are real, and $\varepsilon^2\ll 1$.  
The normal mass hierarchy is realized. 
Because $\gamma$ turns out to be ${\mathcal O}(\varepsilon^2)$, the phases $\rho$ and $\delta$ are calculated from
\begin{eqnarray}
&&
X \left( \propto e^{i\rho}\right)\approx 
\frac{{b\varepsilon ^2 m_0^2 }}{{2\sqrt 2 }}e^{ - i\frac{\alpha }{2}} \cos \frac{\alpha }{2},
\quad
Y \left( \propto e^{-i\delta}\right)\approx 
i\frac{{\sigma b\varepsilon m_0^2 }}{{\sqrt 2 }}e^{ - i\frac{\alpha }{2}} \sin \frac{\alpha }{2},
\label{Eq:Example-XY}
\end{eqnarray}
to be
\begin{eqnarray}
&&
{\rho  \approx  - \frac{\alpha }{2}},
\quad
{\delta  \approx \frac{\alpha }{2} - \frac{\pi }{2}},
\label{Eq:Example-delta-rho}
\end{eqnarray}
for
\begin{eqnarray}
&&
\tan 2\theta _{12} \approx 
\frac{{b\varepsilon ^2 m_0^2 }}{{\sqrt 2  \left( {\Lambda _2  - \Lambda _1 } \right)}}\cos \frac{\alpha }{2},
\quad
\tan 2\theta _{13} \approx  
  \frac{{\sqrt 2 \sigma b\varepsilon m_0^2 }}{{\Lambda _3  - A}}\sin \frac{\alpha }{2},
\label{Eq:Example-12-13}
\end{eqnarray}
from which we find that $X(\propto \varepsilon ^2)$ and $Y(\propto \varepsilon)$ satisfies 
the condition $\left| X\right|\ll\left| Y\right|$. 
From $X$, $\Delta m_ \odot ^2$ is computed to be:
\begin{eqnarray}
&&
\Delta m_ \odot ^2  
\approx \frac{{b\varepsilon ^2 m_0^2 }}{{\sqrt 2 \sin 2\theta _{12} }}\cos \frac{\alpha }{2}.
\label{Eq:Example-sol}
\end{eqnarray}
The dependence of $\varepsilon$ is different for $\tan 2\theta_{12}$ and $\tan 2\theta_{13}$.
As a result, $\tan 2\theta _{12}={\mathcal O}(1)$ is satisfied 
because 
$\Lambda _2  - \Lambda _1= \cos 2\theta_{12}\Delta m^2_\odot$ 
is proportional to $\varepsilon^2 $ 
while $\tan 2\theta _{13}\sim \varepsilon$ is satisfied because of 
$\Lambda _3  - A\approx \Delta m_{atm}^2$ from Eq.(\ref{Eq:Mdagger-M-MassRelation}).

Since the Dirac CP violating phase is given by $\delta + \rho$, we find that $\delta + \rho\approx -\pi/2$. 
This texture shows 
\begin{itemize}
\item almost maximal CP violation 
\end{itemize}
irrespective of the size of $\alpha$. 
The same conclusion can be obtained by evaluating ${\mathcal J}_{CP}$, which is given by
\begin{eqnarray}
&&
{\mathcal J}_{CP} = \frac{{\rm Im}({\rm\bf M}_{e\mu}{\rm\bf M}_{\mu\tau}{\rm\bf M}_{\tau e})}{\Delta m^2_{12}\Delta m^2_{23}\Delta m^2_{31}}
\approx   -\frac{{\sqrt 2 b\varepsilon }}{8}\sin 2\theta _{12} \sin \frac{\alpha }{2},
\label{Eq:Example-J-mass}\\
&&
{\mathcal J}_{CP} = \frac{1}{8}\sin 2\theta _{12} \sin 2\theta _{23} \sin 2\theta _{13} \cos \theta _{13} \sin \delta  
\approx   \frac{{\sqrt 2 b\varepsilon }}{8}\sin 2\theta _{12} \sin \frac{\alpha }{2}\sin \delta, 
\label{Eq:Example-J-angle}
\end{eqnarray}
where we have used $\Delta m^2_{12} = -\Delta m^2_\odot$ and $\Delta m^2_{31} \approx -\Delta m^2_{23} \approx \Delta m^2_{atm}$ 
in Eq.(\ref{Eq:Example-J-mass}), and $\sin 2\theta_{23} \approx \sigma$, $\sin 2\theta_{13}$ from 
Eq.(\ref{Eq:Example-12-13}), and $\cos\theta_{13}\approx 1$ in Eq.(\ref{Eq:Example-J-angle}). 
These two results indicate that $\sin\delta\approx -1$ corresponding to $\delta+\rho \approx -\pi/2$ in 
our estimation, thus confirming 
the conclusion: the emergence of almost maximal CP violation, 
which differs from the conclusion in Ref.\cite{Example}. 
The atmospheric neutrino mixing is almost maximal 
because of $\theta\approx 0$ from $\delta+\rho\approx-\pi/2$ 
and $\phi=0$ from ${\rm\bf M}_{\mu\mu }={\rm\bf M}_{\tau\tau }$ in Eq.(\ref{Eq:Exact-cos23-solution}), 
and is also consistent with Eq.(\ref{Eq:Mdagger-Redinied-23}) giving 
${\rm Im}(e^{-i\rho}{\rm\bf M}_{e\mu})\approx \sigma{\rm Im}(e^{-i\rho}{\rm\bf M}_{e\tau })\approx 2a\varepsilon\sin(\alpha/2)$.
 The advantage of our method lies in the fact that 
the source of the CP violating phase can be directly related to 
${\rm\bf M}_{e\mu,e\tau }$ in $X$ for $\rho$ and in $Y$ for $\delta$.  
The appearance of the almost maximal CP violation can also be read off from the redefined texture given by 
Eq.(\ref{Eq:NuMatrixEntries-1}) with $\alpha _ +=-\alpha _e  =  \rho\approx-\alpha/2$, 
and $\alpha _ -   =  - \gamma \left(\approx 0\right)$:
\begin{eqnarray}
&&
M^\prime_\nu \approx
\frac{{m_0 }}{2}\left( {\begin{array}{*{20}c}
   {a\varepsilon ^2 e^{ - i\alpha } } & {b\varepsilon e^{ - i\left( {\frac{\alpha }{2} - \gamma } \right)} } & { - \sigma b\varepsilon e^{i\left( {\frac{\alpha }{2} - \gamma } \right)} }  \\
   {b\varepsilon e^{ - i\left( {\frac{\alpha }{2} - \gamma } \right)} } & {\left(1+\varepsilon\right)e^{2i\gamma } } & \sigma  \\
   { - \sigma b\varepsilon e^{i\left( {\frac{\alpha }{2} - \gamma } \right)} } & \sigma & {\left(1+\varepsilon\right)e^{ - 2i\gamma }}  \\
\end{array}} \right).
\label{Eq:Example-:Redefined}
\end{eqnarray}
It yields the almost maximal CP violation because this texture approximately satisfies the condition on the maximal CP violation: 
$M^\prime_{e\tau}=-\sigma M^{\prime\ast}_{e\mu}$, $M^\prime_{\tau\tau}=M^{\prime\ast}_{\mu\mu}$, 
 and $M^\prime_{ee}+\sigma M^\prime_{\mu\tau}$=real \cite{Maximal,MassTextureCP1,MassTextureCP10,MassTextureCP11}, 
 where the last condition is valid up to the order $\varepsilon$. 

\section{\label{sec:5}Summary and Discussions}
We have clarified the connection between the CP violating phases and phases of the flavor neutrino masses.  
In terms of the PMNS unitary matrix, any textures can provide $U^{PDG}_{PMNS}$ constructed 
from their three eigenvectors if the flavor neutrinos are appropriately rotated.  However, in terms of 
the flavor neutrino masses $M_\nu$, appropriate rotations modify the phase structure of $M_\nu$ and the 
amount of the rotations can be absorbed in the redefinition of the flavor neutrino masses.  Our formula 
automatically take care of the effect of these rotations, which yield
\begin{eqnarray}
&&
{\rm\bf M}
= \left( \begin{array}{*{20}c}
   A & B & C \\
   B^\ast & D & E   \\
   C^\ast & E^\ast & F\\
\end{array} \right)
\rightarrow
{\rm\bf M}^\prime 
= \left( \begin{array}{*{20}c}
   A & B^\prime & C^\prime \\
   B^{\prime\ast} & D^\prime & E^\prime   \\
   C^{\prime\ast} & E^{\prime\ast} & F^\prime\\
\end{array} \right)
= \left( \begin{array}{*{20}c}
   A & e^{i\left( {\gamma  - \rho } \right)}B & e^{-i\left( {\gamma  + \rho } \right)}C \\
   e^{-i\left( {\gamma  - \rho } \right)}B^\ast & D & e^{ - 2i\gamma }E   \\
   e^{i\left( {\gamma + \rho } \right)}C^\ast & e^{  2i\gamma }E^\ast & F\\
\end{array} \right),
\label{Eq:Mdagger-Redinied-M-U}
\end{eqnarray}
If ${\rm\bf M}$ is employed, the PMNS unitary matrix is $U_{PMNS}$ of Eq.(\ref{Eq:Mdagger-M-PMNS-CP-full}) 
while, if ${\rm\bf M}^\prime$ is employed, it is $U^{PDG}_{PMNS}$ of Eq.(\ref{Eq:U_nu}), where $\rho$ and $\gamma$ 
are absent.  The phases $\rho$ and $\gamma$ are thus redundant. Namely, 
for 
\begin{eqnarray}
&&
\nu^\prime_{flavor} = \left( {\begin{array}{*{20}c}
   e^{-i\rho} & 0 & 0  \\
   0 & e^{-i\gamma} & 0  \\
   0 & 0 & e^{i\gamma}  \\
\end{array}} \right)\nu_{flavor},
\label{Eq:NuRedifinitionPMNS}
\end{eqnarray}
we have found that
\begin{eqnarray}
&&
\nu_{flavor} = U_{PMNS}\left(\delta,\rho,\gamma\right)\nu_{mass}
=U_\nu\left(\delta,\rho,\gamma\right)K\left(\beta_1,\beta_2,\beta_3\right) \nu_{mass}~{\rm for}~{\rm\bf M},
\nonumber\\
&&
\nu^\prime_{flavor}
=
U^{PDG}_{PMNS}\left(\delta + \rho\right) \nu_{mass}
=U_\nu\left(\delta + \rho,0,0\right)K\left(\beta_1-\rho,\beta_2,\beta_3\right) \nu_{mass}~{\rm for}~{\rm\bf M}^\prime.
\label{Eq:SummaryUnitaryMatrix-1}
\end{eqnarray}
For both cases, we obtain ${\mathcal J}_{CP}$ proportional to $\sin(\delta + \rho)$.  It should be 
noted that the Majorana phase for $\nu_1$ is shifted by the rotation of the flavor neutrinos. 
The required amount of the rotations to remove $\rho$ and $\gamma$ in the PMNS unitary matrix 
can be determined from
$\rho = \arg (X)$ with $X =  (c_{23} e^{i\gamma} {\rm\bf M}_{e\mu } - s_{23} e^{-i\gamma} {\rm\bf M}_{e\tau })/c_{13}$,
 and $\gamma = \left(\phi^\prime - \theta^\prime\right)/2$,
where $\theta^\prime$, and $\phi^\prime$ are defined in Eq.(\ref{Eq:Exact-gamma-solution}). 
The Dirac CP phase $\delta$ is determined by  
$\delta = -\arg (Y)$ with $Y =  s_{23} e^{i\gamma} {\rm\bf M}_{e\mu } + c_{23} e^{-i\gamma} {\rm\bf M}_{e\tau }$.
It should emphasized that the phases $\rho$ and $\delta$ are, respectively, associated with 
$\tan 2\theta_{12}$ and $\tan 2\theta_{13}$: 
\begin{eqnarray}
\tan 2\theta_{12}e^{i\rho}\propto X,
\quad
\tan 2\theta_{13}e^{-i\delta}\propto Y.
\label{Eq:Mdagger-Redinied-12-13}
\end{eqnarray}
We have also found that ${\cal J}_{CP}$ is independent of the redefinition of 
masses in Eq.(\ref{Eq:Mdagger-Redinied-M-U}) because of 
${\rm\bf M}^\prime_{e\mu}{\rm\bf M}^\prime_{\mu\tau}{\rm\bf M}^{\prime\ast}_{e\tau}={\rm\bf M}_{e\mu}{\rm\bf M}_{\mu\tau}{\rm\bf M}^\ast_{e\tau}$.

The conditions satisfied by approximately $\mu$-$\tau$ symmetric textures are equivalent to those for textures 
without CP violation, and are described by 
\begin{itemize}
\item $s_{23} e^{i\gamma} {\rm\bf M}_{e\mu } + c_{23} e^{-i\gamma} {\rm\bf M}_{e\tau }\sim 0$,
\item ${\rm\bf M}_{\mu\mu }\sim{\rm\bf M}_{\tau\tau }$.
\end{itemize}
The additional condition is required for textures with CP violation:
\begin{itemize}
\item ${\rm Im}({\rm\bf M}_{\mu\tau }) \sim 0$ corresponding to $\gamma\sim 0$,
\end{itemize}
where $\gamma$ is calculated in Eq.(\ref{Eq:Exact-gamma-solution}). 
If the $\mu$-$\tau$ symmetry breaking parts of ${\rm\bf M}$ have similar strengths, we conclude that  
\begin{itemize}
\item the $\mu$-$\tau$ symmetric $B_+$ is the main source of $\rho$,
\item the $\mu$-$\tau$ symmetry breaking $B_-$ is the main source of $\delta$,
\item the $\mu$-$\tau$ symmetry breaking $D_-$ is the main source of $\cos 2\theta_{23}$,
\item the $\mu$-$\tau$ symmetry breaking $E_-$ is the main source of $\gamma$,
\end{itemize}
where $B_\pm$, $D_-$ and $E_-$ have been defined in Eq.(\ref{Eq:Mnu-mutau-separation-2}). 
As one of the exact results, we have obtained that 
\begin{eqnarray}
\theta_{23} = \pm\pi/4+(\theta  + \phi)/2,
\label{Eq:Exact_23_summary}
\end{eqnarray}
where $\theta$, and $\phi$ are defined in Eq.(\ref{Eq:Exact-cos23-solution}).  
The maximal atmospheric neutrino mixing is realized only if 
$\cos (\delta +\rho)=0$  and ${\rm\bf M}_{\mu\mu } = {\rm\bf M}_{\tau\tau }$ are satisfied. 
Therefore, we observe that 
\begin{itemize}
\item the maximal Dirac CP violation is linked to the maximal atmospheric neutrino mixing,
\end{itemize}
for $ \sin\theta_{13}\neq 0$. This condition is satisfied in the textures discussed 
in Ref.\cite{Maximal,MassTextureCP1,MassTextureCP10,MassTextureCP11}.

To give complete discussions on the leptonic CP violation, we have to consider the Majorana CP 
violation.  As in Eq.(\ref{Eq:SummaryUnitaryMatrix-1}), the redundant phase $\rho$ 
remains in the Majorana phase because Majorana masses directly receive the effect of phases in $M_\nu$.
Since the present method is based on the Hermite matrix $M^\dagger_\nu M_\nu$, the CP 
violating Majorana phases disappear in $M^\dagger_\nu M_\nu$.  To obtain effects of the Majorana phases, 
the simplest way is to  
perform the transformation $U^T_{PMNS} M_\nu U_{PMNS}$=diag.($m_1, m_2, m_3$), 
which is expected to lead to the similar result to the one shown in the Appendix of Ref.\cite{MassTextureCP10}. 
We will discuss this extension of the present method in future publication.

\vspace{3mm}
\noindent
\centerline{\small \bf ACKNOWLEGMENTS}

The work of M.Y. is supported by the Grants-in-Aid for Scientific Research on Priority Areas from the 
Ministry of Education, Culture, Sports, Science, and Technology, Japan.

\appendix
\section{\label{sec:Appendix1}$\mu$-$\tau$ Symmetric Texture and Breaking Part}
In this Appendix, we show the parameterization of our neutrino mass matrix.  
For a given $M_\nu$, we can formally 
divide $M_\nu$ into the $\mu$-$\tau$ symmetric part $M_{sym}$ and its 
breaking part $M_b$ expressed in terms of 
$M^{(\pm)}_{e\mu} = (M_{e\mu} \pm (-\sigma M_{e\tau}))/2$ ($\sigma=\pm 1$) and 
$M^{(\pm)}_{\mu\mu} = (M_{\mu\mu} \pm M_{\tau\tau})/2$ \cite{MassTextureCP1,MassTextureCP10}:
\begin{eqnarray}
&&
M_\nu =  \left( \begin{array}{*{20}c}
   M_{ee} & M_{e\mu } & - \sigma M_{e\mu }  \\
   M_{e\mu } & M_{\mu\mu } & M_{\mu \tau }   \\
    - \sigma M_{e\mu } & M_{\mu \tau } & M_{\mu\mu }\\
\end{array} \right)
=
M_{sym} + M_b
\label{Eq:Mnu-mutau-separation}
\end{eqnarray}
with
\begin{eqnarray}
&&
M_{sym}  = \left( \begin{array}{*{20}c}
   M_{ee} & M^{(+)}_{e\mu } & - \sigma M^{(+)}_{e\mu }  \\
   M^{(+)}_{e\mu } & M^{(+)}_{\mu\mu } & M_{\mu \tau }   \\
    - \sigma M^{(+)}_{e\mu } & M_{\mu \tau } & M^{(+)}_{\mu\mu }\\
\end{array} \right),
\quad
M_b  = \left( \begin{array}{*{20}c}
   0 & M^{(-)}_{e\mu } & \sigma M^{(-)}_{e\mu }  \\
   M^{(-)}_{e\mu }& M^{(-)}_{\mu\mu } & 0  \\
   \sigma M^{(-)}_{e\mu } & 0 & - M^{(-)}_{\mu\mu } \\
\end{array} \right).
\label{Eq:Mnu-mutau-separation-2}
\end{eqnarray}
The lagrangian for $M_{sym}$: $-{\mathcal{L}}_{mass}=\nu^TM_{sym}\nu/2$ with 
$\nu=(\nu_e, \nu_\mu, \nu_\tau)^T$ turns out to be invariant under the exchange of 
$\nu_\mu\leftrightarrow -\sigma\nu_\tau$.  We parameterize ${\rm\bf M}(=M^\dagger_\nu M_\nu)$ as ${\rm\bf M}$=
${\rm\bf M}_{sym}$+${\rm\bf M}_b$:
\begin{eqnarray}
&&
{\rm\bf M}_{sym}  = \left( \begin{array}{*{20}c}
   A & B_+ & - \sigma B_+  \\
   B^\ast_+ & D_+ & E_+   \\
   - \sigma B^\ast_+ & E_+ & D_+\\
\end{array} \right),
\quad
{\rm\bf M}_b  = \left( \begin{array}{*{20}c}
   0 & B_- & \sigma B_-  \\
   B^\ast_- & D_- & iE_-  \\
   \sigma B^\ast_- & -iE_- & - D_-\\
\end{array} \right).
\label{Eq:Mdagger-M}
\end{eqnarray}
where
\begin{eqnarray}
&&
A = \left| {M_{ee} } \right|^2  + 2\left( {\left| {M_{e\mu }^{( + )} } \right|^2  + \left| {M_{e\mu }
^{( - )} } \right|^2 } \right) ,
\nonumber \\ 
&&
B_ +   = M_{ee}^ \ast  M_{e\mu }^{( + )}  + M_{e\mu }^{( + ) \ast } \left( {M_{\mu \mu }^{( + )}  - 
\sigma M_{\mu \tau } } \right) + M_{e\mu }^{( - ) \ast } M_{\mu \mu }^{( - )}  ,
\nonumber \\ 
&&
B_ -   = M_{ee}^ \ast  M_{e\mu }^{( - )}  + M_{e\mu }^{( - ) \ast } \left( {M_{\mu \mu }^{( + )}  + 
\sigma M_{\mu \tau } } \right) + M_{e\mu }^{( + ) \ast } M_{\mu \mu }^{( - )},
\nonumber \\ 
&&
D_ +   = \left| {M_{e\mu }^{( + )} } \right|^2  + \left| {M_{e\mu }^{( - )} } \right|^2  + \left| {M_
{\mu \mu }^{( + )} } \right|^2  + \left| {M_{\mu \mu }^{( - )} } \right|^2  + \left| {M_{\mu \tau } }
 \right|^2 ,
\nonumber \\ 
&&
D_ -   = 2{\rm Re} \left( {M_{e\mu }^{( - ) \ast } M_{e\mu }^{( + )}  + M_{\mu \mu }^{( - ) \ast } M_
{\mu \mu }^{( + )} } \right),
\nonumber \\ 
&&
E_+ = {\rm Re}(E) = \sigma \left( {\left| {M_{e\mu }^{( - )} } \right|^2  - \left| {M_{e\mu }^{( + )} } \right|^2 }
 \right) + 2{\rm Re} \left( {M_{\mu \mu }^{( + ) \ast } M_{\mu \tau } } \right),
\nonumber\\
&&
E_- = {\rm Im}(E) = 2{\rm Im} \left( {M_{\mu \mu }^{( - ) \ast } M_{\mu \tau }  - \sigma M_{e\mu }^{( - ) \ast } M_
{e\mu }^{( + )} } \right),
\label{Eq:A-F}
\end{eqnarray}
for $E=E_++iE_-$. Similarly, we define 
$B$=$B_++B_-$, $C$=$-\sigma (B_+-B_-)$, $D$=$D_++D_-$, and $F$=$D_+-D_-$ to describe matrix elements of ${\rm\bf M}$. 

By the redefinition of the flavor neutrino $\nu_{flavor}=(\nu_e, \nu_\mu, \nu_\tau)^T$ as
\begin{eqnarray}
&&
\nu^\prime_{flavor} = \left( {\begin{array}{*{20}c}
   e^{i\alpha_e} & 0 & 0  \\
   0 & e^{i\alpha_\mu} & 0  \\
   0 & 0 & e^{i\alpha_\tau}  \\
\end{array}} \right)\nu_{flavor}
=
e^{i\alpha_e}\left( {\begin{array}{*{20}c}
  1 & 0 & 0  \\
   0 & e^{i\left(\alpha_+ + \alpha_-\right)} & 0  \\
   0 & 0 & e^{i\left(\alpha_+ - \alpha_-\right)}  \\
\end{array}} \right)\nu_{flavor},
\label{Eq:NuRedifinition}
\end{eqnarray}
for $\alpha_+=(\alpha_\mu+\alpha_\tau)/2-\alpha_e$ and $\alpha_-=(\alpha_\mu-\alpha_\tau)/2$, 
this $M_\nu$ becomes equivalent to
\begin{eqnarray}
&&
M^\prime_\nu = e^{-2i\alpha_e } \left( {\begin{array}{*{20}c}
   {M_{ee} } & {e^{-i\left( {\alpha _ +   + \alpha _ -  } \right)} M_{e\mu } } & {e^{-i\left( {\alpha _ +   - \alpha _ -  } \right)} M_{e\tau } }  \\
   {e^{-i\left( {\alpha _ +   + \alpha _ -  } \right)} M_{e\mu } } & {e^{-2i\left( {\alpha _ +   + \alpha _ -  } \right)} M_{\mu \mu } } & {e^{-2i\alpha _ +  } M_{\mu \tau } }  \\
   {e^{-i\left( {\alpha _ +   - \alpha _ -  } \right)} M_{e\tau } } & {e^{-2i\alpha _ +  } M_{\mu \tau } } & {e^{-2i\left( {\alpha _ +   - \alpha _ -  } \right)} M_{\tau \tau } }  \\
\end{array}} \right).
\label{Eq:NuMatrixEntries-1}
\end{eqnarray}
The relevant interactions are 
kept invariant as 
$g{\overline \ell}\gamma^\mu W^{(-)}_\mu \nu_{flavor}/\sqrt 2-\nu_{flavor}^T M_\nu\nu_{flavor}/2=g{\overline \ell^\prime}\gamma^\mu W^{(-)}_\mu \nu_{flavor}^\prime/\sqrt 2-\nu_{flavor}^{\prime T} M^\prime_\nu\nu_{flavor}^\prime/2$, 
where $\ell =(\ell_e, \ell_\mu,\ell_\tau)^T\equiv (e, \mu, \tau)^T $ and $\ell^\prime_f = e^{i\alpha_f}\ell_f$ for $f=e,\mu,\tau$.
Similarly, 
\begin{eqnarray}
&&
{\rm\bf M}^\prime = \left( {\begin{array}{*{20}c}
   A & e^{-i\left( {\alpha _ +   + \alpha _ - } \right)} B & e^{-i\left( {\alpha _ +   - \alpha _ - } \right)} C   \\
   e^{ i\left( {\alpha _ +   + \alpha _ -  } \right)} B^\ast & D & e^{ 2i\alpha _ -  } E  \\
   e^{ i\left( {\alpha _ +   - \alpha _ -  } \right)} C^\ast & e^{-2i\alpha _ -  } E^\ast  & F  \\
\end{array}} \right),
\label{Eq:NuMatrixEntries-2}
\end{eqnarray}
is equivalent to ${\rm\bf M}$ in Eq.(\ref{Eq:Mdagger-M}). 

It is instructive to note that there are three typical forms of 
the PMNS unitary matrix depending on how the flavor neutrinos are redefined:
\begin{enumerate}
\item $U_{PMNS}$ with $\delta$, $\rho$ and $\gamma$
\begin{eqnarray}
&&
 \left( {\begin{array}{*{20}c}
   1 & 0 & 0  \\
   0 & {e^{i\gamma } } & 0  \\
   0 & 0 & {e^{ - i\gamma } }  \\
\end{array}} \right)\left( {\begin{array}{*{20}c}
   {c_{12} c_{13} } & {s_{12} c_{13} e^{i\rho } } & {s_{13} e^{ - i\delta } }  \\
   { - c_{23} s_{12} e^{ - i\rho }  - s_{23} c_{12} s_{13} e^{i\delta } } & {c_{23} c_{12}  - s_{23} s_{12} s_{13} e^{i\left( {\delta  + \rho } \right)} } & {s_{23} c_{13} }  \\
   {s_{23} s_{12} e^{ - i\rho }  - c_{23} c_{12} s_{13} e^{i\delta } } & { - s_{23} c_{12}  - c_{23} s_{12} s_{13} e^{i\left( {\delta  + \rho } \right)} } & {c_{23} c_{13} }  \\
\end{array}} \right)
\nonumber\\
&&
\cdot\left( {\begin{array}{*{20}c}
   {e^{i\beta _1 } } & 0 & 0  \\
   0 & {e^{i\beta _2 } } & 0  \\
   0 & 0 & {e^{i\beta _3 } }  \\
\end{array}} \right),
\label{Eq:PMNSMatrices1}
\end{eqnarray}
for
\begin{eqnarray}
&&
\left( {\begin{array}{*{20}c}
   A & B & C  \\
   {B^ *  } & D & E  \\
   {C^ *  } & {E^ *  } & F  \\
\end{array}} \right),
\end{eqnarray}
\item $U_{PMNS}$ with $\delta$ and $\rho$
\begin{eqnarray}
&&
 \left( {\begin{array}{*{20}c}
   {c_{12} c_{13} }  \\
   { - c_{23} s_{12} e^{ - i\rho }  - s_{23} c_{12} s_{13} e^{i\delta } }  \\
   {s_{23} s_{12} e^{ - i\rho }  - c_{23} c_{12} s_{13} e^{i\delta } }  \\
\end{array}{\rm{ }}\begin{array}{*{20}c}
   {s_{12} c_{13} e^{i\rho } }  \\
   {c_{23} c_{12}  - s_{23} s_{12} s_{13} e^{i\left( {\delta  + \rho } \right)} }  \\
   { - s_{23} c_{12}  - c_{23} s_{12} s_{13} e^{i\left( {\delta  + \rho } \right)} }  \\
\end{array}{\rm{ }}\begin{array}{*{20}c}
   {s_{13} e^{ - i\delta } }  \\
   {s_{23} c_{13} }  \\
   {c_{23} c_{13} }  \\
\end{array}} \right)\left( {\begin{array}{*{20}c}
   {e^{i\beta _1 } } & 0 & 0  \\
   0 & {e^{i\beta _2 } } & 0  \\
   0 & 0 & {e^{i\beta _3 } }  \\
\end{array}} \right),
\label{Eq:PMNSMatrices2}
\end{eqnarray}
for
\begin{eqnarray}
&&
\left( {\begin{array}{*{20}c}
   A & {e^{i\gamma } B} & {e^{ - i\gamma } C}  \\
   {e^{ - i\gamma } B^ *  } & D & {e^{ - 2i\gamma } E}  \\
   {e^{i\gamma } C^ *  } & {e^{2i\gamma } E^ *  } & F  \\
\end{array}} \right),
\end{eqnarray}
\item $U_{PMNS}$ with $\delta^\prime(=\delta+\rho)$
\begin{eqnarray}
&&
 \left( {\begin{array}{*{20}c}
   {c_{12} c_{13} }  \\
   { - c_{23} s_{12}  - s_{23} c_{12} s_{13} e^{i \delta^\prime} }  \\
   {s_{23} s_{12}  - c_{23} c_{12} s_{13} e^{i\delta^\prime} }  \\
\end{array}{\rm{ }}\begin{array}{*{20}c}
   {s_{12} c_{13} }  \\
   {c_{23} c_{12}  - s_{23} s_{12} s_{13} e^{i\delta^\prime} }  \\
   { - s_{23} c_{12}  - c_{23} s_{12} s_{13} e^{i\delta^\prime} }  \\
\end{array}{\rm{ }}\begin{array}{*{20}c}
   {s_{13} e^{ - i\delta^\prime} }  \\
   {s_{23} c_{13} }  \\
   {c_{23} c_{13} }  \\
\end{array}} \right)\left( {\begin{array}{*{20}c}
   {e^{i\left( {\beta _1  - \rho } \right)} } & 0 & 0  \\
   0 & {e^{i\beta _2 } } & 0  \\
   0 & 0 & {e^{i\beta _3 } }  \\
\end{array}} \right),
\label{Eq:PMNSMatrices3}
\end{eqnarray}
for
\begin{eqnarray}
&&
 \left( {\begin{array}{*{20}c}
   A & {e^{ - i\left( {\rho  - \gamma } \right)} B} & {e^{ - i\left( {\rho  + \gamma } \right)} C}  \\
   {e^{i\left( {\rho  - \gamma } \right)} B^ *  } & D & {e^{ - 2i\gamma } E}  \\
   {e^{i\left( {\rho  + \gamma } \right)} C^ *  } & {e^{2i\gamma } E^ *  } & F  \\
\end{array}} \right).
\end{eqnarray}
\end{enumerate}

\section{\label{sec:Appendix2}Formula for Masses, Mixings and Phases.}
By adopting $U_{PMNS}$ of Eq.(\ref{Eq:Mdagger-M-PMNS-CP-full}) to diagonalize ${\rm\bf M}$,  
which $U^\dagger_{PMNS}{\rm\bf M}U_{PMNS}$=diag.($m^2_1, m^2_2, m^2_3$) is satisfied. 
We, then, obtain that
\begin{eqnarray}
&&
\tan 2\theta _{12}e^{i\rho}  = \frac{{2X}}{{\Lambda _2  - \Lambda_1 }},
\quad
\tan 2\theta _{13}e^{-i\delta}  = \frac{2Y}{{\Lambda _3  - A}},
\label{Eq:Mdagger-ExactMixingAngles12-13}\\ 
&&
{\rm Re} \left( {e^{ - 2i\gamma } E} \right)\cos 2\theta _{23}  + D_-\sin 2\theta _{23}
  + i{\rm Im} \left( {e^{ - 2i\gamma } E} \right) =  - s_{13} e^{- i\left(\delta+\rho\right)} \left( e^{-i\rho} X\right)^\ast,
\label{Eq:Mdagger-ExactMixingAngles23}
\end{eqnarray}
for three mixing angles and three phases, and
\begin{eqnarray}
&&
m_1^2  = c_{12}^2 \Lambda _1  + s_{12}^2 \Lambda _2  - 2c_{12} s_{12} e^{-i\rho} X,
\quad
m_2^2  = s_{12}^2 \Lambda _1  + c_{12}^2 \Lambda _2  + 2c_{12} s_{12} e^{-i\rho} X,
\quad
m_3^2  = \frac{{c_{13}^2 \Lambda _3  - s_{13}^2 A}}{{c_{13}^2  - s_{13}^2 }}, 
\label{Eq:Mdagger-ExactMasses}
\end{eqnarray}
for three masses, where
\begin{eqnarray}
&&
X = \frac{{c_{23} e^{i\gamma} B - s_{23} e^{-i\gamma} C}}{c_{13} }, 
\quad
Y = s_{23} e^{i\gamma } B + c_{23} e^{ - i\gamma } C,
\nonumber\\
&&
\Lambda _1  = \frac{{c_{13}^2 A - s_{13}^2 \Lambda _3 }}{{c_{13}^2  - s_{13}^2 }},
\quad
\Lambda _2  = c_{23}^2 D + s_{23}^2 F - 2s_{23} c_{23} {\rm Re} \left( {e^{ - 2i\gamma } E} \right),
\quad
\nonumber\\
&&
\Lambda _3  = s_{23}^2 D + c_{23}^2 F + 2s_{23} c_{23} {\rm Re} \left( {e^{ - 2i\gamma } E} \right).
\label{Eq:Mdagger-MassParameters}
\end{eqnarray}
Note that the Dirac CP violation involves the angle $\delta + \rho$.
There are useful relations:
\begin{eqnarray}
&&
A \approx 
\frac{{\Sigma m_ \odot ^2  - \cos 2\theta _{12} \Delta m_ \odot ^2  + s_{13}^2 \left( {2\Delta m_{atm}^2  + \cos 2\theta _{12} \Delta m_ \odot ^2 } \right)}}{2},
\nonumber\\
&&
D_ +   \approx \frac{1}{2}\left( {\Delta m_{atm}^2  + \sum m_ \odot ^2  + \frac{{\cos 2\theta _{12} \Delta m_ \odot ^2  - s_{13}^2 \left( {2\Delta m_{atm}^2  + \cos 2\theta _{12} \Delta m_ \odot ^2 } \right)}}{2}} \right),
\nonumber\\
&&
{\rm Re}\left(e^{ - 2i\gamma }  E \right) - 2D_ -  \Delta  \approx \frac{1}{2}\sigma \left( {\Delta m_{atm}^2  - \frac{{\cos 2\theta _{12} \Delta m_ \odot ^2  + s_{13}^2 \left( {2\Delta m_{atm}^2  + \cos 2\theta _{12} \Delta m_ \odot ^2 } \right)}}{2}} \right),
\nonumber\\
&&
\Lambda _1  =
\frac{{\Sigma m_ \odot ^2  - \cos 2\theta _{12} \Delta m_ \odot ^2 }}{2},
\quad
\Lambda _2  = 
\frac{{\Sigma m_ \odot ^2  + \cos 2\theta _{12} \Delta m_ \odot ^2 }}{2},
\nonumber\\
&&
\Lambda _3  \approx \frac{{2\Delta m_{atm}^2  + \Sigma m_ \odot ^2  - s_{13}^2 \left( {2\Delta m_{atm}^2  + \cos 2\theta _{12} \Delta m_ \odot ^2 } \right)}}{2},
\label{Eq:Mdagger-M-MassRelation}
\end{eqnarray}
up to ${\mathcal{O}}(s^2_{13})$, where 
$\sum m_ \odot ^2  = m_1^2  + m_2^2$ and $\vert\Delta m_ {atm}^2\vert\gg\Delta m_ \odot ^2$ is used to neglect other terms.  The term $\tan 2\theta_{12}$ in Eq.(\ref{Eq:Mdagger-ExactMixingAngles12-13}) is 
identically satisfied by these $\Lambda_{1,2}$ because
\begin{eqnarray}
&&
X= \frac{1}{2}e^{i\rho}\sin 2\theta_{12}\Delta m^2_\odot,
\label{Eq:Mdagger-M-X}
\end{eqnarray}
is obtained from Eq.(\ref{Eq:Mdagger-ExactMasses}).


\newpage



\begin{thebibliography}{}
\bibitem{Sun}
	Y. Fukuda {\it et al.}, [Super-Kamiokande Collaboration], \Journal{\PRL}{81}{1158}{1998}; [\Journal{\Erratum}{81}{4297}{1998}];
	B.T. Clevel {\it et al.}, \Journal{\APJ}{496}{505}{1998};
	W. Hampel {\it et al.}, [GNO Collaboration],  \Journal{\PLB}{447}{127}{1999};
	Q.A. Ahmed. {\it et al.}, [SNO Collaboration], \Journal{\PRL}{87}{071301}{2001}; \Journal{\PRL}{89}{011301}{2002}.
See also,
    J.N. Bahcall, W.A. Fowler, I. Iben and R.L. Sears, \Journal{\APJ}{137}{344}{1963};
    J. Bahcall, \Journal{\PRL}{12}{300}{1964};
    R. Davis, Jr., \Journal{\PRL}{12}{303}{1964};
    R. Davis, Jr., D.S. Harmer and K.C. Hoffman, \Journal{\PRL}{20}{1205}{1968}; 
    J.N. Bahcall, N.A. Bahcall and G. Shaviv, \Journal{\PRL}{20}{1209}{1968};
    J.N. Bahcall and R. Davis, Jr., \Journal{\SCI}{191}{264}{1976}.

\bibitem{K2K}
	S. H. Ahn, {\it et al.}, [K2K Collaboration], \Journal{\PLB}{511}{178}{2001}; \Journal{\PRL}{90}{041801}{2003}.

\bibitem{Reactor} 
	M. Apollonio, {\it et al.}, [CHOOZ Collaboration], \Journal{\EPJ}{27}{331}{2003};
	K. Eguchi, {\it et al.}, [KamLAND collaboration], \Journal{\PRL}{90}{021802}{2003}.
	K. Inoue, [KamLAND collaboration], \Journal{\NJP}{6}{147}{2004};

\bibitem{Experiments}
	Y. Suzuki, ``Accelerator and Atmospheric Neutrinos", Talk given at {\it XXII International Symposium on Lepton-Photon Interactions at High Energy}, Uppsala, Sweden (June 30-July 5, 2005);
	A. Poon, ``Solar and Reactor Neutrinos", Talk given at {\it XXII International Symposium on Lepton-Photon Interactions at High Energy}, Uppsala, Sweden (June 30-July 5, 2005);

\bibitem{SK}
	Y. Fukuda {\it et al.}, [Super-Kamiokande Collaboration], \Journal{\PRL}{81}{1562}{1998}; \Journal{\PRL}{82}{2430}{1999};
	T. Kajita for the collaboration, \Journal{\NPSUPPL}{77}{123}{1999};
	A. Suzuki, ``Evidence for Neutrino Mass", 
	Talk given at {\it Neutrino 2006: The XXII International Conference on Neutrino Physics and Astrophysics}, 
	Santa Fe, New Mexico, USA (June 13-19, 2006).
	See also,
	T. Kajita and Y. Totsuka, \Journal{\RMP}{73}{85}{2001}.

\bibitem{CPProposal}
	For a recent review, O. Mena, \Journal{\MPL}{20}{1}{2005};
	``Getting the most from Long Baseline Neutrino Experiments, Theoretical Overview", 
	Talk given at {\it Neutrino 2006: The XXII International Conference on Neutrino Physics and Astrophysics}, 
	Santa Fe, New Mexico, USA (June 13-19, 2006);
	W.J. Marciano, ``Neutrino Superbeams and Leptonic CP Violation", 
	Talk given at {\it Neutrino 2006: The XXII International Conference on Neutrino Physics and Astrophysics}, 
	Santa Fe, New Mexico, USA (June 13-19, 2006);
	L. Camilleri, ``Low energy Superbeams, Beta beams and the Neutrino Factory", 
	Talk given at {\it Neutrino 2006: The XXII International Conference on Neutrino Physics and Astrophysics}, 
	Santa Fe, New Mexico, USA (June 13-19, 2006).

\bibitem{CPViolationOrg}
	S.M. Bilenky, J. Hosek and S.T. Petcov, \Journal{\PLBOLD}{94B}{495}{1980};
	J. Schechter and J.W.F. Valle, \Journal{\PRD}{22}{2227}{1980};
	M. Doi, T. Kotani, H. Nishiura, K. Okuda and E. Takasugi, \Journal{\PLBOLD}{102B}{323}{1981}.

\bibitem{PMNS} 
	B. Pontecorvo, \Journal{\JETPUSSR}{7}{172}{1958} [\Journal{\ZETP}{34}{247}{1958}];
	Z. Maki, M. Nakagawa and S. Sakata, \Journal{\PTP}{28}{870}{1962}. 

\bibitem{Jarlskog}
	C. Jarlskog, \Journal{\PRL}{55}{1039}{1985}.

\bibitem{NuData} 
	G.L. Fogli, E. Lisi, A. Marrone, A. Palazzo, \Journal{\PPNP}{57}{742}{2006}.
See also,
	S. Goswami, A. Bandyopadhyay and S. Choubey, \Journal{\NPSUPPL}{143}{121}{2005};
	G. Altarelli, \Journal{\NPSUPPL}{143}{470}{2005};
	A. Bandyopadhyay, \Journal{\PLB}{608}{115}{2005}.

\bibitem{leptogenesis} 
	M. Fukugida and T. Yanagida, \Journal{\PLBOLD}{174}{45}{1986}.

\bibitem{CP-Baryon} 
	A.D. Sakharov, \Journal{\JETPUSSRLETT}{5}{24}{1967} [\Journal{\PismaZETP}{5}{32}{1967}].

\bibitem{PDG} 
	S. Eidelman et al. (Particle Data Group), \Journal{\PLB}{592}{149}{2004}.
	See also, L.-L. Chau and W.-Y. Keung, \Journal{\PRL}{53}{1802}{1984}.

\bibitem{JarlskogMass} 
	G. C. Branco, T. Morozumi, B.M. Nobre and M.N. Rebelo. \Journal{\NPB}{617}{475}{2001}; 
	G. C. Branco, R. Gonzalez Felipe, F. R. Joaquim, I.Masina, M. N. Rebelo and C. A. Savoy, 
	\Journal{\PRD}{67}{073025}{2003}.

\bibitem{Nishiura}
	T. Fukuyama and H. Nishiura, in {\it Proceedings of International Workshop on Masses and Mixings of Quarks and Leptons} edited by Y. Koide (World Scientific, Singapore, 1997), p.252; ``Mass Matrix of Majorana Neutrinos", [arXiv:hep-ph/9702253];
	Y. Koide, H. Nishiura, K. Matsuda, T. Kikuchi and T. Fukuyama, \Journal{\PRD}{66}{093006}{2002};
	Y. Koide, \Journal{\PRD}{69}{093001}{2004};
	K. Matsuda and H. Nishiura, \Journal{\PRD}{69}{117302}{2004}; \Journal{\PRD}{71}{073001}{2005}; \Journal{\PRD}{72}{033011}{2005};  \Journal{\PRD}{73}{013008}{2006}. 

\bibitem{mu-tau}
	R.N. Mohaptra and S. Nussinov,\Journal{\PRD}{60}{013002}{1999};
	Z.Z. Xing, \Journal{\PRD}{61}{057301}{2000}; \Journal{\PRD}{64}{093013}{2001};
	C.S. Lam, \Journal{\PLB}{507}{214}{2001}; \Journal{\PRD}{71}{093001}{2005};
	``Mass Independent Textures and Symmetry", [arXiv: hep-ph/0611017];
	E. Ma and M. Raidal, \Journal{\PRL}{87}{011802}{2001}; [\Journal{\Erratum}{87}{159901}{2001}];
	A. Datta and P.J. O'Donnell, \Journal{\PRD}{72}{113002}{2005};

\bibitem{mu-tau0}
	W. Grimus and L. Lavoura,  \Journal{\JHEP}{0107}{045}{2001}; \Journal{\EPJ}{28}{123}{2003}; \Journal{\PLB}{572}{189}{2003}; \Journal{\PLB}{579}{113}{2004}; \Journal{\JPG}{30}{1073}{2004}; \Journal{\JHEP}{0508}{013}{2005};
	W. Grimus, A.S. Joshipura, S. Kaneko, L. Lavoura and M. Tanimoto, \Journal{\JHEP}{0407}{078}{2004}; 
	W. Grimus, A.S. Joshipura, S. Kaneko, L. Lavoura, H. Sawanaka and M. Tanimoto, \Journal{\NPB}{713}{151}{2005}. 
	W. Grimus, S. Kaneko, L. Lavoura, H. Sawanaka and M. Tanimoto, \Journal{\JHEP}{0601}{110}{2006}. 

\bibitem{mu-tau1}
	T. Kitabayashi and M. Yasu\`{e}, \Journal{\PLB}{524}{308}{2002}; 
	\Journal{\IJMP}{17}{2519}{2002}; \Journal{\PRD}{67}{015006}{2003};
	I. Aizawa, M. Ishiguro, T. Kitabayashi and M. Yasu\`{e}, \Journal{\PRD}{70}{015011}{2004};
	I. Aizawa, T. Kitabayashi and M. Yasu\`{e}, \Journal{\PRD}{71}{075011}{2005}.

\bibitem{mu-tau2}
	R.N. Mohapatra,  \Journal{\JHEP}{0410}{027}{2004};
	R.N. Mohapatra and S. Nasri,  \Journal{\PRD}{71}{033001}{2005};
	R.N. Mohapatra, S. Nasri and Hai-Bo Yu, \Journal{\PLB}{615}{231}{2005}; \Journal{\PRD}{72}{033007}{2005}; \Journal{\PLB}{636}{114}{2006}; \Journal{\PLB}{639}{318}{2006}.

\bibitem{MassTextureCP1} 
	T. Kitabayashi and M. Yasu\`{e}, \Journal{\PLB}{621}{133}{2005}.

\bibitem{MassTextureCP10} 
	I. Aizawa and M. Yasu\`{e}, \Journal{\PRD}{73}{015002}{2006}.

\bibitem{MassTextureCP11} 
	I. Aizawa, T. Kitabayashi and M. Yasu\`{e}, \Journal{\PRD}{72}{055014}{2005}; \Journal{\NPB}{728}{220}{2005}.

\bibitem{MassTextureCP2} 
See for example,
	F. Plentinger and W. Rodejohann, \Journal{\PLB}{625}{264}{2005};
	R.N. Mohapatra and W. Rodejohann, \Journal{\PRD}{72}{053001}{2005}; 
	W. Grimus, S. Kaneko, L. Lavoura, H. Sawanaka and M. Tanimoto, \Journal{\JHEP}{0601}{110}{2006} in Ref.\cite{mu-tau1}; 
	Z.Z. Xing, \Journal{\PRD}{74}{013009}{2006}; \Journal{\PLB}{641}{189}{2006};
	Z.Z. Xing, H. Zhang, and S. Zhou, \Journal{\PLB}{641}{189}{2006}; 
	Y.H. Ahn, S.K. Kang, C.S. Kim and J. Lee, \Journal{\PRD}{73}{093005}{2006};
	``$\mu$-$\tau$ Symmetry and Radiatively Generated Leptogenesis", [arXiv: hep-ph/0610007];
	Z.Z. Xing and S. Zhou, ``TeV-scale Leptogenesis and Tri-bimaximal Neutrino Mixing in the Minimal Seesaw Model", [arXiv: hep-ph/0607302];
	U. Sarkar and S.K. Singh, ``CP Violation in Neutrino Mass Matrix", [arXiv: hep-ph/0608030];
	G.C. Branco, R. G. Felipe, F.R. Joaquim, ``A New Bridge between Leptonic CP Violation and Leptogenesis", [arXiv: hep-ph/0609297];
	R.N. Mohapatra, Hai-Bo Yu, \Journal{\PLB}{644}{346}{2007};
	Y. Farzana and A.Yu. Smirnov, \Journal{\JHEP}{0701}{059}{2007};
	S.K. Agarwalla, M.K. Parida, R.N. Mohapatra, G. Rajasekaran,``Neutrino Mixings and Leptonic CP Violation from CKM Matrix and Majorana Phases", [arXiv: hep-ph/0611225] (to appear in Phys. Rev. D (2007));
	S. Pascoli, S.T. Petcov and A. Riotto, ``Leptogenesis and Low Energy CP Violation in Neutrino Physics", [arXiv: hep-ph/0611338];
	S. Luo and Z.Z. Xing, ``Friedberg-Lee Symmetry Breaking and Its Prediction for $\theta_{13}$", [arXiv: hep-ph//0611360] (to appear in Phys. Lett. B (2007)).

\bibitem{Fuki}
	K. Fuki and M. Yasu\`{e}, \Journal{\PRD}{73}{055014}{2006}.

\bibitem{Maximal}
	E. Ma, \Journal{\MPL}{17}{2361}{2002};
	K.S. Babu, E. Ma and J.W.F. Valle, \Journal{\PLB}{552}{207}{2003};
	W. Grimus and L. Lavoura, \Journal{\PLB}{579}{113}{2004} in Ref.\cite{mu-tau1}.

\bibitem{naturalness}
    G. 't Hooft, in {\it Recent Development in Gauge Theories: Proceedings of the Cargese Summer Institute} edited by G. 't Hooft et al. (Plenum Press, New York, 1980), p.135 (NATO Advanced Study Institutes Series: Series B, Physics, Vol 59).

\bibitem{leptonic-mu-tau}
	See for example, 
	W. Grimus and L. Lavoura,  \Journal{\JHEP}{0107}{045}{2001}, \Journal{\EPJ}{28}{123}{2003}, and \Journal{\JPG}{30}{73}{2004} in Ref.\cite{mu-tau1}; 
	``A Three-Parameter Model for the Neutrino Mass Matrix", [arXiv: hep-ph/0611149];
	E. Ma and G. Rajasekaran, \Journal{\PRD}{68}{071302}{2003};
	R.N. Mohapatra, in Ref.\cite{mu-tau2};
	K. Matsuda and H. Nishiura, in Ref.\cite{Nishiura};	
	W. Grimus, A.S. Joshipura, S. Kaneko, L. Lavoura, H. Sawanaka and M. Tanimoto, in Ref.\cite{mu-tau1};
	W. Grimus, ``Realization of $\mu$-$\tau$ Interchange Symmetry", [arXiv: hep-ph/0610158], 
	Talk given at {\it ICHEP06: The XXXIII International Conference on High Energy Physics}, 
	Moscow, Russia (July 26 - August 2, 2006);
	E. Ma, ``Application of Finite Groups to Neutrino Mass Matrices, [arXiv: hep-ph/0612013]; 
	and other references therein; ``Lepton Family Symmetry and Possible Application to the Koide Mass Formula", 
	[arXiv: hep-ph/0612022].

\bibitem{recent_mu-tau-breaking}
For recent studies, see for example,
	R.N. Mohapatra, S. Nasri and Hai-Bo Yu, \Journal{\PLB}{636}{114}{2006} in Ref.\cite{mu-tau2}; 
	N. Haba and W. Rodejohann, \Journal{\PRD}{74}{017701}{2006}.

\bibitem{YasueTalk}
	M. Yasu\`{e}, ``Complex Neutrino Mass Matrix and PMNS Unitary Matrix" (unpublished), 
	Talk given at {\it 2005 Shizuoka Workshop on Quark and Lepton Masses and Mixings}, 
	University of Shizuoka, Shizuoka, Japan (Dec. 20-21, 2005).

\bibitem{EigenVectors}
	C.S. Lam, \Journal{\PRD}{74}{113004}{2006}.

\bibitem{GeneralCP} 
	I. Aizawa and M. Yasu\`{e}, \Journal{\PLB}{607}{267}{2005}.

\bibitem{Example} 
	R.N. Mohapatra, S. Nasri and Hai-Bo Yu, \Journal{\PRD}{72}{033007}{2005} in Ref.\cite{mu-tau2}.
\end{thebibliography}
\end{document}